\documentclass[aps,prd]{revtex4-2}
\usepackage{dcolumn}
\usepackage{bm}
\usepackage{bbm}
\usepackage{morefloats}
\usepackage{color}
\usepackage{amsmath}
\usepackage{mathrsfs}
\usepackage{slashed}
\usepackage{multirow}
\usepackage{amssymb}
\usepackage{bbding}
\usepackage{appendix}
\usepackage{multirow}
\usepackage{graphicx}
\usepackage{caption}
\usepackage{subfigure}
\usepackage{float}
\usepackage{indentfirst}  
\usepackage{xcolor}
\usepackage{verbatim}
\usepackage[marginal]{footmisc}
\usepackage{ulem} 

\makeatletter
\newcommand*{\rom}[1]{\expandafter\@slowromancap\romannumeral #1@}
\makeatother

\begin{document}

\title{ Resolving the polarization puzzles in $D^0\to VV$ }

\author{Ye Cao$^{1,2}$\footnote{Email: caoye@ihep.ac.cn}, Yin Cheng$^{1,2}$\footnote{Email: chengyin@ihep.ac.cn}, Qiang Zhao$^{1,2}$\footnote{E-mail: zhaoq@ihep.ac.cn} }

\affiliation{ 1) Institute of High Energy Physics,
        Chinese Academy of Sciences, Beijing 100049, P.R. China}

\affiliation{ 2) University of Chinese Academy of Sciences, Beijing 100049, P.R. China}

\begin{abstract}
  We carry out a systematic analysis of the Cabbibo-favored (CF) and singly Cabbibo-suppressed (SCS) decays of $D^0\to VV$, and demonstrate that the long-distance mechanism due to the final-state interactions (FSIs) can provide a natural explanation for these mysterious polarization puzzles observed in $D^0\to VV$ in experiments. More observables, which can be measured at BESIII and possibly at LHCb, are also suggested.
\end{abstract}

\maketitle

\section{introduction}

In the past two decades, ``polarization puzzles'' arose from the decays of heavy mesons into two vectors. In the beauty sector, naive power counting predicts that $B \to VV$ ($V=\phi, \ K^*, \ \rho$, and $\omega$) decays are dominated by the longitudinal polarization since the transverse polarization amplitudes suffer from the helicity-flipping suppression at the order of $\Lambda_{\text{QCD}}/m_b$. This prediction is confirmed in $B^0 \to \rho^+\rho^-$, $B^+ \to \rho^0\rho^+$ and $\rho^0 K^{*+}$~\cite{BaBar:2003zor, BaBar:2003rdc, Belle:2003lsm}, which has indicated the helicity conservation~\cite{Brodsky:1981kj,Chernyak:1981zz,Chernyak:1983ej} in $B\to VV$. However, apparent deviations were found in $B \to \phi K^*$, where the longitudinal polarization only accounts for about $50\%$ of the decay rate~\cite{BaBar:2003zor, Belle:2003ike}.

In the charm sector, the situation is more complicated since the heavy quark expansion method becomes unreliable here. Although different phenomenological approaches and techniques have been developed in the literature, such as the flavor SU(3) symmetry model~\cite{Kamal:1990ky}, broken flavor SU(3) symmetry model~\cite{Hinchliffe:1995hz}, pole-dominance model~\cite{Bedaque:1993fb}, factorization approach~\cite{Bauer:1986bm,Kamal:1990ky,Cheng:2010rv,Jiang:2017zwr,Uppal:1992se}, and the heavy quark effective Lagrangian and chiral perturbation theory~\cite{Bajc:1997ey}, and they describe well some of the $D\to VV$ decay channels, a systematic and coherent study of all the Cabibbo-favored (CF) and singly Cabibbo-suppressed (SCS) decays is still unavailable.

The naive factorization model~\cite{ElHassanElAaoud:1999min} and the Lorentz invariant-based symmetry model~\cite{Hiller:2013cza} indeed predict that the longitudinal polarization fraction (defined as $f_L$) may not be dominant in $D\to VV$. However, the predictions seem to have quantitatively deviated from the experimental measurements. Meanwhile, experimental measurements reveal unexpected puzzling results which cannot be explained by theory. For instance, the MARK-III measurement of $D^0\to \bar{K}^{*0}\rho^0$ shows the dominance of the transverse polarization~\cite{MARK-III:1991fvi}, though it suffers from a large uncertainty. In contrast, the precise measurement of $D^0\to \rho^0\rho^0$ by the FOCUS Collaboration shows large longitudinal polarization fractions of $f_L=(71\pm 4\pm 2)\%$~\cite{FOCUS:2007ern}. Recently, the angular distribution of $D^0\to \omega\phi$ is measured by the BESIII collaboration. It is stunning to find that the final states $\omega$ and $\phi$ seem to be fully transversely polarized with $f_L=0.00\pm 0.10 \pm 0.08$, which corresponds to $f_L < 0.24$ at $95\%$ confidence level~\cite{BESIII:2021raf}. In contrast, the partial wave measurement by CLEO-c~\cite{dArgent:2017gzv} shows that the decay of $D^0\to \phi\rho^0$ is dominated by the $S$-wave with $BR(D^0\to (\phi\rho^0)_{S-\text{wave}})=(1.40\pm 0.12)\times 10^{-3}$. The total b.r. $BR(D^0\to(\phi\omega)\simeq BR(D^0\to(\phi\omega)_T)=(0.65\pm 0.10)\times 10^{-3}$ from BESIII~\cite{BESIII:2021raf} turns out to be much smaller than that of $D^0\to \phi\rho^0$ from CLEO-c, i.e. $BR(D^0\to \phi\rho^0)=(1.56\pm 0.13)\times 10^{-3}$ ~\cite{dArgent:2017gzv}. Such a significant difference is also confirmed by LHCb~\cite{LHCb:2018mzv}. These puzzling observations show that, although the decay of $D^0\to VV$ has been one of the broadly studied processes, we still lack of knowledge about some crucial pieces of dynamics in its decay mechanisms.

In this work, apart from the leading short-distance mechanisms, i.e. the color-allowed direct emission (DE), color-suppressed (CS) internal $W$-emission, and the color-suppressed flavor internal conversion (IC) by the $W$-exchange between the quark and anti-quark inside $D^0$, we propose that the non-factorizable final state interactions (FSIs) as a long-distance dynamics could be a key to resolving the mysterious ``polarization puzzle" in $D^0\to VV$. Notice that for $D\to VV$ the threshold of $VV$ is not far below the $D$ meson mass. Namely, the charm quark is not heavy enough. The presence of the near-threshold vector-meson rescatterings either in an $S$ or $P$ wave may introduce significant long-distance dynamics. It indicates the necessity for a proper treatment of the non-factorizable FSIs in $D\to VV$. It thus motivates us to carry out a systematic and coherent investigation of $D^0\to VV$ with the long-distance FSIs taken into account.

As follows, we first analyze the leading short-distance mechanisms where the DE and CS couplings will be calculated in a non-relativistic constituent quark model (NRCQM). An effective Lagrangian approach (ELA) will be adopted for deriving the FSI transition amplitudes. We introduce the FSIs to those CF and SCS exclusive $VV$ channels. Numerical results and discussions will be presented in the end with a brief summary.

\section{framework}
With the kinematic constraint, $D^0$ can access 14 $VV$ decay channels, i.e. 3 CF channels: $K^{*-}\rho^+$, $\bar{K}^{*0}\rho^0$, $\bar{K}^{*0}\omega$; 8 SCS channels: $K^{*+}K^{*-}$, $K^{*0}\bar{K}^{*0}$, $\rho^+\rho^-$, $\rho^0\rho^0$, $\omega\omega$, $\rho^0\omega$, $\rho^0\phi$, $\omega\phi$; and 3 doubly-Cabibbo-suppressed (DSC) channels: $K^{*+}\rho^-$, $K^{*0}\rho^0$, $K^{*0}\omega$. To quantify the transition mechanisms, we distinguish between the short-distance dynamics for the quark-level transitions in the quark model and the long-distance dynamics arising from the hadronic level interactions.

\subsection{Short-distance mechanisms in the quark model}
\begin{figure}[H]
  \centering
  \includegraphics[width=3.3cm]{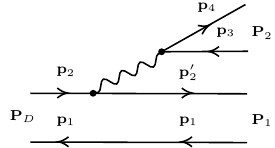}
  \includegraphics[width=3.3cm]{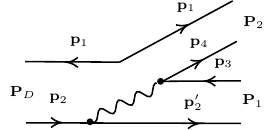}
  \includegraphics[width=3.3cm]{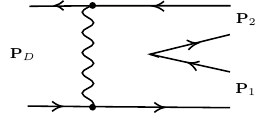}
  \includegraphics[width=3.3cm]{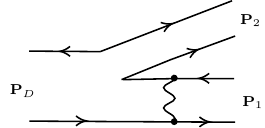}
  \includegraphics[width=3.6cm]{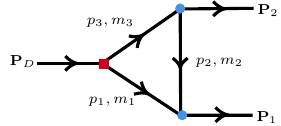}
  \caption{Schematic diagrams for the $D^0\to VV$ decays via the short and long-distance transition mechanisms. (a), (b), (c) and (d) denote the short-distance transitions and stand for the DE, CS, and IC-processes ((c) and (d)), respectively; (e) illustrates the hadron-level FSIs. } 
    \label{fig: tree diagrams}
\end{figure}

The typical leading order transition mechanisms via the DE, CS and IC-processes are illustrated in Fig.~\ref{fig: tree diagrams} (a)-(d), respectively. Apart from the weak coupling strengths, these three processes correspond to three topologically distinguishable amplitudes from the short-distance dynamics. In the SU(3) flavor symmetry limit and given the dominance of the short-distance dynamics, all the $D^0\to VV$ decays can be described by linear combinations of these three leading amplitudes.

\subsubsection{``$1\to 3$" transitions in the quark model}
The DE and CS transitions involve the process of ``$1\to 3$" decays of the initial charm quark. Their corresponding coupling strengths $g_{\text{DE}}^{(\text{P})}$ and $g_{\text{CS}}^{(\text{P})}$ are defined by the transition matrix element as follows:
\begin{eqnarray}
    i{\cal M}_{\text{DE/CS}}^{(\text{P})} &= &\langle V_1({\bf P}_1;J_1, J_{1z})V_2({\bf P}_2;J_2, J_{2z})|\hat{H}_{W,1\to 3}^{(\text{P})}|D^0({\bf P}_D;J_i, J_{iz})\rangle \nonumber\\
     &\equiv&g_{\text{DE/CS}}^{(\text{P});J_{iz};J_{1z},J_{2z}}V_{cq}V_{uq}\ ,
\end{eqnarray}
where ${\bf P}_1={\bf p}_1+{\bf p}_2'$ and ${\bf P}_2={\bf p}_3+{\bf p}_4$ are the momentum conservation relations for the DE transition, and ${\bf P}_1={\bf p}_1+{\bf p}_4$ and ${\bf P}_2={\bf p}_2'+{\bf p}_3$ are for the CS one; $\hat{H}_{W,1\to 3}^{(\text{P})}$ is the operator which takes different forms for the parity-violated (PV) or parity-conserved (PC) transitions, and has been derived in Refs.~\cite{LeYaouanc:1988fx, Richard:2016hac, Niu:2020gjw}. The above formula contains spatial wavefunction integrals for which the NRCQM wavefunctions~\cite{Kokoski:1985is,Godfrey:1985xj,Godfrey:1986wj} are adopted. In Appendix~\ref{appendix:1} we present the detailed expressions for the transition amplitudes of ``$1\to 3$". 

It can be seen that the mass differences within those $VV$ channels will lead to different values for both $g_{\text{DE}}^{(\text{P})}$ and $g_{\text{CS}}^{(\text{P})}$ after taking the wavefunction convolutions as an additional source of the SU(3) flavor symmetry breaking.

\begin{table}
  \caption{Amplitudes of all the CF and SCS decay channels for $D^0\to VV$ via the short-distance dynamics. The upper and lower panels are for the CF and SCS processes, respectively. }
  \label{table:short-dist-amp}
  \begin{tabular}{|c|c|}
      \hline\hline
      Decay channels & Amplitudes \\  
      \hline
      $K^{*-}\rho^+$ & $[g_{\text{DE}}^{(\text{P})}+e^{i\theta}g_{\text{IC}(s\bar{d})}^{(\text{P})}]V_{cs}V_{ud}$ \\
      $\bar{K}^{*0}\rho^0$ & $\frac{1}{\sqrt{2}}[g_{\text{CS}}^{(\text{P})}-e^{i\theta}g_{\text{IC}(s\bar{d})}^{(\text{P})}]V_{cs}V_{ud}$ \\
      $\bar{K}^{*0}\omega$ & $\frac{1}{\sqrt{2}}[g_{\text{CS}}^{(\text{P})}+e^{i\theta}g_{\text{IC}(s\bar{d})}^{(\text{P})}]V_{cs}V_{ud}$ \\
      \hline
      $K^{*+}K^{*-}$ & $[g_{\text{DE}}^{(\text{P})}+e^{i\theta}g_{\text{IC}(s\bar{s})}^{(\text{P})}]V_{cs}V_{us}$ \\
      $K^{*0}\bar{K}^{*0}$ & $e^{i\theta}[g_{\text{IC}(s\bar{s})}^{(\text{P})}V_{cs}V_{us}+g_{\text{IC}(d\bar{d})}^{(\text{P})}V_{cd}V_{ud}]$ \\
      $\rho^+\rho^-$ & $[g_{\text{DE}}^{(\text{P})}+e^{i\theta} g_{\text{IC}(d\bar{d})}^{(\text{P})}]V_{cd}V_{ud}$ \\
      $\rho^0\rho^0$ & $\frac12 [-g_{\text{CS}}^{(\text{P})} + e^{i\theta}g_{\text{IC}(d\bar{d})}^{(\text{P})}]V_{cd}V_{ud}$ \\
      $\omega\omega$ & $\frac12 [g_{\text{CS}}^{(\text{P})} + e^{i\theta}g_{\text{IC}(d\bar{d})}^{(\text{P})}]V_{cd}V_{ud}$ \\
      $\rho^0\omega$ & $-\frac12 e^{i\theta}g_{\text{IC}(d\bar{d})}^{(\text{P})}V_{cd}V_{ud}$\\
      $\phi\rho^0$ & $\frac{1}{\sqrt{2}}g_{\text{CS}}^{(\text{P})}V_{cs}V_{us}$ \\
      $\phi\omega$ & $\frac{1}{\sqrt{2}}g_{\text{CS}}^{(\text{P})}V_{cs}V_{us}$ \\
      \hline
  \end{tabular}
\end{table}

\subsubsection{``$2\to 2$" transitions in the quark model}
The IC transitions involve the process of ``$2\to 2$" scatterings of the initial charm and $\bar{u}$ into the CF $\bar{K}^* V$ channel or SCS channels.  The corresponding coupling strength is defined as $g_{\text{IC}(q\bar{q}')}^{(\text{P})}$. As shown by Fig.~\ref{fig: tree diagrams}(c) and (d), different intermediate $q\bar{q}$ poles may contribute. It implies that this mechanism involves significant SU(3) flavor symmetry breakings. From the three CF transitions of $D^0\to \bar{K}^* V$ and the experimental data for these three channels, we see the relation, $|g_{\text{DE}}^{(\text{P})}|>|g_{\text{CS}}^{(\text{P})}|>|g_{\text{IC}(s\bar{d})}^{(\text{P})}|$. 

The nonlocal operators for the weak and strong transitions will distinguish processes between Figs.~\ref{fig: tree diagrams} (c) and (d). The weak matrix element and the strong matrix element are connected by the intermediate resonance state as propagators. By separately calculating the weak internal conversion and strong transition couplings, Figs.~\ref{fig: tree diagrams} (c) and (d) together
will be evaluated explicitly in the quark model, provided that the wavefunctions of the initial, final and intermediate states can be described stably. In the following, we will refer to both Figs.~\ref{fig: tree diagrams} (c) and (d) as IC processes. Intuitively, the amplitudes of these two Ic processes may cancel due to the opposite sign of the real part of the propagators. Meanwhile, the IC transition deserves some attention in the $D$ meson decays due to the relatively light mass of the $D$ meson. In Fig.~\ref{fig: tree diagrams} the transition process (d) is often treated as dual with (b). However, such a treatment should not be true since Fig.~\ref{fig: tree diagrams} (d) involves a pole structure in the transition amplitude while (b) does not. Moreover, we also categorize Fig.~\ref{fig: tree diagrams} (c) as an IC transition since it also involves a pole structure in the amplitude. The difference between Fig.~\ref{fig: tree diagrams} (c) and (d) is that the weak interaction occurs before or after the strong quark-pair creation, and these two coupling vertices are connected by intermediate propagators. A general expression for the sum of Fig.~\ref{fig: tree diagrams} (c) and (d) can be written as
\begin{align}
    \begin{split}
        i{\cal M}_{\text{IC}}^{\text{(P)}}&=\sum_{M_n(0^{\pm})}\langle V_1({\bf P}_1;J_{1z})V_2({\bf P}_2;J_{2z})|\hat{H}_S|M_n({\bf P}_D;J_{iz})\rangle \frac{i}{m_D^2-m_{M_n}^2+im_{M_n}\Gamma_{M_n}}\langle M_n({\bf P}_D;J_{iz}) |\hat{H}_{W,2\to 2}^{(\text{P})}|D^0({\bf P}_D;J_{iz})\rangle\\
        &+\sum_{D_n^*(1^{\pm})}\langle V_1({\bf P}_1;J_{1z})|H_{W,2\to2}^{(\text{P})}| D_n^*({\bf P}_1;J_{1z})\rangle\frac{i}{m_{V_1}^2-m_{D_n^*}^2+im_{D_n^*}\Gamma_{D_n^*}} \langle D_n^*({\bf P}_1;J_{1z})V_2({\bf P}_2;J_{2z})|\hat{H}_{S}^{(\text{P})}|D^0({\bf P}_D;J_{iz})\rangle\\
        & \equiv g_{\text{IC}}^{(\text{P});J_{iz};J_{1z},J_{2z}}V_{cq}V_{uq}\ ,
    \end{split}
\end{align}
where cancellation between (c) and (d) is apparent. One also sees that if the initial mass $m_D>> m_i^{(c)}$ and $m_i^{(d)}>>m_V$, the IC contributions will be suppressed by the propagators. Then, as expected, Fig.~\ref{fig: tree diagrams} (b) will be more predominant and (c) and (d). For the case of $D^0\to VV$, the initial $D$ meson mass is actually not heavy enough. It means that the transition of Fig.~\ref{fig: tree diagrams} (c) with the intermediate excited states may even be enhanced instead of suppressed by the propagator~\cite{Niu:2021qcc}, while in Fig.~\ref{fig: tree diagrams} (d) the propagator suppression is rather clear since the mass of the intermediate charmed masses are even more heavier than the initial $D$ meson. Such a situation suggests that the cancellation between Fig.~\ref{fig: tree diagrams} (c) and (d) may not be as complete as in the case of e.g. $B\to VV$ where the QCD factorization works better.

In Table~\ref{table:short-dist-amp},  we collect the short-distance amplitudes of all the CF and SCS channels. With (P)=(PC) or (PV) there are actually six quantities to account for the PC and PV transitions in each channel. It should be pointed out that in the SU(3) flavor symmetry limit, these quantities will take the same values in all the $D^0\to VV$ channels. However, with the consequence of the SU(3) flavor symmetry breaking, they can be different. In fact, although the DE transition is dominant, contributions from the CS and IC transitions cannot be neglected. Therefore, a more realistic evaluation of these quantities beyond the simple parametrization is necessary. Considering the complexity of $g_{\text{IC}(q\bar{q}')}^{(\text{P})}$ and its relatively small magnitudes, we leave it to be determined by the combined analysis.

One notices that an additional phase angle $\theta$ is introduced in Tab.~\ref{table:short-dist-amp}. By calculating the DE and CS in the NRCQM, these two terms can be determined with a fixed phase. $\theta$ describes the relative phase between the IC and the DE/CS amplitudes and we mention in advance that $\theta=180^{\circ}$ is favored.

\subsection{Long-distance mechanisms via the FSIs}
In order to clarify the role played by the long-distance transition mechanisms, we start with the analysis of the SCS processes $D^0\to \phi\rho^0$ and $\phi\omega$. Regarding the leading short-distance transitions, one finds that only the CS process (Fig.~\ref{fig: tree diagrams}(b)) can contribute. In addition, note that $\rho^0$ and $\omega$ are produced by the $u\bar{u}$ component and they are degenerate in mass. These two unique features imply that given the dominance of the short-distance mechanism, these two channels should have the same decay rates. The isospin decomposition gives,
    $u\bar{u}=\frac12(u\bar{u}+d\bar{d})+\frac12(u\bar{u}-d\bar{d})=\frac{1}{\sqrt{2}}(|\omega\rangle+|\rho^0\rangle)$,
where $|\omega\rangle$ and $|\rho^0\rangle$ correspond to the flavor wavefunctions of $\omega$ and $\rho^0$, respectively. Thus, the coupling strength of the CS transition for $D^0\to \phi\rho^0$ and $D^0\to \phi\omega$ can be expressed as
\begin{eqnarray}
    &&i{\cal M}_{(\text{P})}(D^0\to\phi\rho^0/\phi\omega) \nonumber\\
    &=&\langle \phi\rho^0/\phi\omega|\phi (u\bar{u})\rangle\langle \phi(u\bar{u})|H_{W(\text{P})}^{(\text{CS})}|D^0\rangle\nonumber\\
    &=&\frac{1}{\sqrt{2}}g_{\text{CS}}^{(\text{P})}V_{cs}V_{us} \ ,
\end{eqnarray}
where (P) in the above equation can be either (PV) or (PC) denoting the amplitudes for the PV or PC transitions. It actually shows that with the leading order approximation, these two decays should have the same decay rate and the same polarization behavior. However, the partial wave measurement by Ref.~\cite{dArgent:2017gzv} shows that the decay of $D^0\to \phi\rho^0$ is dominated by the $S$-wave with $BR(D^0\to (\phi\rho^0)_{S-\text{wave}})=(1.40\pm 0.12)\times 10^{-3}$ and its total b.r. is $BR(D^0\to \phi\rho^0)=(1.56\pm 0.13)\times 10^{-3}$. The $S$-wave can only come from the parity-violated transition. The relatively small $P$-wave contribution indicates the relatively small contributions from the parity-conserved mechanism. In contrast, the recent measurement of $D^0\to \phi\omega$ by BESIII shows that this channel is dominated by the transverse polarization, i.e. $BR(D^0\to(\phi\omega)_T)=(0.65\pm 0.10)\times 10^{-3}$. Surprisingly, its b.r. of the longitudinally polarized decay is negligibly small. Although these two measurements involve two different observables, the suppression of the longitudinal polarization contributions and the significant difference of their total b.r. suggests that these two decay channels involve mechanisms beyond the leading short-distance transitions.

Recognizing that the mass of $K^*\bar{K}^*$ is almost degenerate to those of $\phi\rho^0/\phi\omega$ and the decays of $D^0\to K^{*+}K^{*-}$ and $D^0\to K^{*0}\bar{K}^{*0}$ actually involve different processes in
Fig.~\ref{fig: tree diagrams}, we anticipate that the decays of $D^0\to \phi\rho^0$ and $D^0\to \phi\omega$ should acquire different contributions from the intermediate $K^{*+}K^{*-}$ and $K^{*0}\bar{K}^{*0}$ rescatterings to the isovector channel $\phi\rho^0$ and isoscalar channel $\phi\omega$, respectively. Generally speaking, intermediate processes which have sizeable b.r.s into  $\phi\rho^0$ and $\phi\omega$ may contribute as long-distance mechanisms as illustrated in Fig.~\ref{fig: tree diagrams}(e). However, taking into account the mass thresholds and weak coupling strengths, only some of those $PP$, $VP$ and $VV$ channels can have sizeable effects.

In Tab.~\ref{weak-coup} we list the processes which contains the DE transitions as the leading contributing channels to the FSIs and we adopt their DE couplings extracted in the NRCQM in the loop calculation. This is a reasonable approximation since they are the dominant processes for $D^0\to VV$. The data will be fitted with all the mechanisms included. Interestingly, one sees that the intermediate $PP$ and $VP$ channels contribute to the $VV$ channels differently due to the parity constraint. This allows us to extract the weak couplings from the available data for the $PP$ and $VP$. In contrast, the weak couplings for $D^0\to VV$ contains both PC and PV components. 

For the decays of $D^0\to \phi\rho^0$ and $D^0\to \phi\omega$ we only consider the rescatterings of $D^0\to K^{*+}K^{*-}\to \phi\rho^0$ and $\phi\omega$ as the leading long-distance amplitudes, but neglect contributions from the CS processes. Note that although $D^0\to K^{*+}K^{*-}$ and $\rho^+\rho^-$ are the DE processes, their experimental measurements are still unavailable. We will extract the DE amplitudes in the NRCQM as the theoretical input. For the decays of $D^0\to VV$ the kinematics and local weak coupling operators make it a reliable estimate of the DE and CS transition amplitudes~\cite{Niu:2020gjw}.

\begin{table}[H]
  \centering
  \caption{The weak couplings of CF and SCS channels in units of $10^{-6}$ which are estimated by calculating the DE-process in the NRCQM and the uncertainty comes from the model parameters for $VV$ modes and extracted by matching the experimental data for $PP$ and $VP$ modes.}
  \label{weak-coup}
  \footnotesize
  \begin{tabular}{cccc}
      \hline\hline
      $VV$ Modes            &b.r. of DE                         &$g_{\text{DE}}^{(\text{PC})}[\text{GeV}^{-1}]$            &$g_{\text{DE}}^{(\text{PV})}[\text{GeV}]$\\
      \hline
      ${K^{*-}\rho^+}$   &${0.22\pm 0.06}$                  &${2.61\pm 0.43}$                         &${4.90\pm 0.63}$\\
      ${K^{*+}K^{*-}}$   &${(0.89\pm 0.29)\%}$              &${2.91\pm 0.60}$                         &${5.45\pm 0.89}$\\
      ${\rho^+\rho^-}$   &${(1.33\pm 0.49)\%}$              &${2.96\pm 0.88}$                         &${4.74\pm 0.80}$\\                 
      \hline\hline
      $PP/VP$ Modes                 &b.r. of Expt.                        &$g^{(\text{PC})}$                         &$g^{(\text{PV})}[\text{GeV}]$\\
      \hline
      $K^-\pi^+$            &$(3.95\pm0.03)\%$                    &$0$                                &$2.64\pm 0.01$\\
      $K^+K^-$              &$(4.08\pm0.06)\times 10^{-3}$        &$0$                                &$3.84\pm 0.03$\\
      $\pi^+\pi^-$          &$(1.45\pm0.02)\times 10^{-3}$        &$0$                                &$2.19\pm 0.02$\\
      \hline
      $K^{*-}\pi^+$         &$(6.93\pm1.20)\%$                    &$1.29\pm 0.11$                     &$0$\\
      $\rho^+K^-$           &$(11.20\pm0.70)\%$                   &$1.54\pm 0.05$                     &$0$\\
      $K^{*-}K^+$           &$(1.86\pm0.30)\times 10^{-3}$        &$1.16\pm 0.09$                     &$0$\\
      $K^{*+}K^-$           &$(5.67\pm0.90)\times 10^{-3}$        &$2.02\pm 0.16$                     &$0$\\
      $\rho^-\pi^+$         &$(5.15\pm0.25)\times 10^{-3}$        &$1.23\pm 0.03$                     &$0$\\
      $\rho^+\pi^-$         &$(1.01\pm0.04)\%$                    &$1.72\pm 0.03$                     &$0$\\
      \hline       
  \end{tabular}
\end{table}

One also notices that the decays of $D^0\to\phi\rho^0$ and $\phi\omega$ actually receive different interfering contributions from the intermediate $K^{*+}K^{*-}$ rescatterings. Namely, the DE transition can access both channels, while the IC transition only contributes to the $\phi\omega$ channel. It means that these two channels will receive different interfering contributions from the intermediate $K^{*+}K^{*-}$. To illustrate this explicitly, we write down the leading $K^{*+}K^{*-}$ rescattering amplitudes through triangle loops by exchanging $\mathbb{K}$ ($K$ or $K^*$) respectively as follows:
\begin{eqnarray}
i\mathcal{M}^{loop}_{(\text{P})\phi\rho^0}
&=&\frac{1}{\sqrt{2}}g_{\text{DE}}^{(\text{P})}V_{cs}V_{us}\sum_{(\mathbb{K})}\tilde{\mathcal{I}}[(\text{P});K^{*+},K^{*-},(\mathbb{K})] \ , \\
i\mathcal{M}^{loop}_{(\text{P})\phi\omega}
&=&\left(\frac{1}{\sqrt{2}}g_{\text{DE}}^{(\text{P})}+e^{i\theta}g_{\text{IC}(s\bar{s})}^{(\text{P})}\right)V_{cs}V_{us}\nonumber\\
&&\times \sum_{(\mathbb{K})}\tilde{\mathcal{I}}[(\text{P});K^{*+},K^{*-},(\mathbb{K})] \ ,
\end{eqnarray}
where the sum is over the contributing meson loops $\tilde{\mathcal{I}}[(\text{P});K^{*+},K^{*-},(\mathbb{K})]$; as defined before, (P) (=(PC) or (PV)) indicates the PC or PV property of the corresponding amplitude. The triangle loop function $\tilde{\mathcal{I}}$ has different integrand functions for different loops. 

Taking the PC loop transition $[(\text{PC});K^{*},\bar{K}^{*},(K)]$ as an example, the loop integral is:
\begin{eqnarray}
        &&\tilde{\mathcal{I}}[(\text{PC});K^{*+},K^{*-},(K)]\nonumber\\
       & =&\int\frac{d^4p_1}{(2\pi)^4}V_{1\mu\nu}D^{\mu\mu^{\prime}}(K^{*})V_{2\mu^{\prime}}D(K)V_{3\nu^{\prime}}D^{\nu\nu^{\prime}}(\bar{K}^{*})\mathcal{F}(p_i^2),\label{eq: im1}
    \end{eqnarray}
where the vertex functions have compact forms as follows:
\begin{align}
    \begin{split}
        V_{1\mu\nu}&=-i\epsilon_{\alpha\beta\mu\nu}p_1^{\alpha}p_3^{\beta},\\
        V_{2\mu^{\prime}}&=ig_{V_1 K^*\bar{K}}\epsilon_{\alpha_1\beta_1\mu^{\prime}\delta}p_1^{\alpha_1}p_{V_1}^{\beta_1}\varepsilon_{V_1}^{\delta*},\\
        V_{3\nu^{\prime}}&=ig_{V_2 \bar{K}^*K}\epsilon_{\alpha_2\beta_2\nu^{\prime}\lambda}p_3^{\alpha_2}p_{V_2}^{\beta_2}\varepsilon_{V_2}^{\lambda*} \ ,
        \label{eq: vertex-1P}
    \end{split}
\end{align}
with $V_1$ and $V_2$ denoting the final state $\phi$ and $\rho^0/\omega$, respectively. In Eq.~(\ref{eq: im1}) functions $D^{\mu\mu^{\prime}}(K^*)={-i(g^{\mu\mu^{\prime}}-{p^{\mu}p^{\mu^{\prime}}}/{p^2})}/(p^2-m_{K^*}^2+i\epsilon)$ and $D(K)={i}/(p^2-m_K^2+i\epsilon)$ are the propagators for $K^*$ and $K$, respectively, with four-vector momentum $p$. We note that all the vertex couplings involving the light pseudoscalar ($P$) and vector ($V$) meson couplings, i.e. $g_{VPP}$, $g_{VVP}$, and $g_{VVV}$, have been extracted by Refs.~\cite{Cheng:2021nal,Cheng:2023lov}, such as $g_{V_1 K^*\bar{K}}$ and $g_{V_2 \bar{K}^*K}$ in Eq.~(\ref{eq: vertex-1P}). In Appendix~\ref{appendix:2} the detailed integrals for the relevant loop transitions have been given. 

In order to cut off the ultra-violet (UV) divergence in the loop integrals, a commonly-adopted form factor is included to regularize the integrand:
\begin{eqnarray}
    \mathcal{F}(p_i^2)=\prod_i(\frac{\Lambda_i^2-m_i^2}{\Lambda_i^2-p_i^2}),
\end{eqnarray}
where $\Lambda_i\equiv m_i+\alpha\Lambda_{\text{QCD}}$ with $m_i$ the mass of the $i$-th internal particle, and $\Lambda_{\text{QCD}}=220$ MeV with $\alpha=1\sim 2$ as the cut-off parameter~\cite{Guo:2010ak}.

\section{results and discussions}

\subsection{Fitting scheme}
In our approach there are limited numbers of parameters to be fitted by the available data. Apart from the phase angle $\theta$ and cut-off parameter $\alpha$, the IC couplings,  i.e. $g_{\text{IC}(s\bar{d})}^{(\text{P})}$, $g_{\text{IC}(s\bar{s})}^{(\text{P})}$, $g_{\text{IC}(d\bar{d})}^{(\text{P})}$, are treated as free parameters and will be determined by the overall fitting. Concerning the phase angle $\theta$ our numerical study shows that $\theta=180^\circ$ is favored. This indicates a natural phase between the short and long-distance amplitudes. Namely, a sign may arise from between the quark-level and hadronic level amplitudes due to the convention adopted. Also, the results seem not to be sensitive to $\alpha$ within a reasonable range of values. Hence, we first restrict $\alpha=1.4\pm 0.14$ as an overall parameter and then fit the IC couplings to the existing experimental data.

Note that the IC coupling $g_{\text{IC}(s\bar{d})}^{(\text{P})}$ appears in the CF modes, i.e. $D^0\to K^{*-}\rho^+$, $\bar{K}^{*0}\rho^0$, and $\bar{K}^{*0}\omega$, while $g_{\text{IC}(s\bar{s})}^{(\text{P})}$ and $g_{\text{IC}(d\bar{d})}^{(\text{P})}$ appear in the SCS modes, i.e. $D^0\to K^{*+}K^{*-}$, $K^{*0}\bar{K}^{*0}$, $\rho^+\rho^-$, $\rho^0\rho^0$, $\omega\omega$, and $\rho^0\omega$. Experimental measurements of the CF  decays of $D^0\to K^{*-}\rho^+$, $\bar{K}^{*0}\rho^0$, and $\bar{K}^{*0}\omega$ can be found in the literature~\cite{ARGUS:1992gpk,dArgent:2017gzv,BESIII:2017jyh,MARK-III:1991fvi}. The SCS decays of $K^{*0}\bar{K}^{*0}$~\cite{dArgent:2017gzv} and $\rho^0\rho^0$~\cite{dArgent:2017gzv,FOCUS:2007ern} were also measured by experiment. However, one notices that there are quite significant differences between the results from Refs.~\cite{dArgent:2017gzv} and \cite{FOCUS:2007ern}. 

To proceed, we adopt the data for $D^0\to K^{*-}\rho^+$, $\bar{K}^{*0}\rho^0$, $\bar{K}^{*0}\omega$, and $K^{*0}\bar{K}^{*0}$ as input for the determination of the IC couplings. The fitting results for these input channels are listed in Tab.~\ref{IC-coupling-fit} as a comparison. The numerical study shows that in the SCS transitions $g_{\text{IC}(s\bar{s})}^{(\text{PC})}\simeq g_{\text{IC}(d\bar{d})}^{(\text{PC})}\simeq (1.0\sim 1.2) \times 10^{-6} \  \ \text{GeV}^{-1}$ and $g_{\text{IC}(s\bar{s})}^{(\text{PV})}\simeq g_{\text{IC}(d\bar{d})}^{(\text{PV})}\simeq  (0.8\sim 1.0)\times10^{-6} \ \text{GeV}$ can be determined. We also find that the coupling $g_{\text{IC}(s\bar{d})}^{(\text{P})}$ in the CF transition is different from $g_{\text{IC}(s\bar{s})}^{(\text{P})}$
in the SCS, i.e. $g_{\text{IC}(s\bar{d})}^{(\text{PC})}=(0.2\sim 0.5)\times 10^{-6} \ \text{GeV}^{-1}$ and $g_{\text{IC}(s\bar{d})}^{(\text{PV})}=(2.5\sim3.0)\times 10^{-6} \ \text{GeV}$.
This is understandable since these quantities describe different intermediate flavor configurations in the IC transitions which can also be associated with SU(3) flavor symmetry breaking. With these fitted quantities, we can then calculate the polarization and partial-wave b.r.s of all the other CF and SCS channels as the predictions of our model. In particular, a comparison with the measured channels of $D^0\to \rho^0\rho^0$~\cite{dArgent:2017gzv,FOCUS:2007ern}, $\phi\rho^0$~\cite{dArgent:2017gzv} and $\phi\omega$~\cite{BESIII:2021raf} can serve as a test of our model.

\begin{table}[H]
    \centering
    \caption{The fitted branching ratios in comparison with the experimental data in our framework. The best fitting gives $g_{\text{IC}(s\bar{d})}^{\text{(PC)}}\simeq(0.2\sim 0.5)\times 10^{-6} \ \text{GeV}^{-1}$, $g_{\text{IC}(s\bar{d})}^{(\text{PV})}\simeq(2.5\sim 3.0)\times 10^{-6} \ \text{GeV}$, $g_{\text{IC}(s\bar{s})}^{\text{(PC)}}\simeq g_{\text{IC}(d\bar{d})}^{\text{(PC)}}\simeq(1.0\sim 1.2)\times 10^{-6} \ \text{GeV}^{-1}$, and $g_{\text{IC}(s\bar{s})}^{\text{(PV)}}\simeq g_{\text{IC}(d\bar{d})}^{\text{(PV)}}\simeq(0.8\sim 1.0)\times 10^{-6} \ \text{GeV}$.}\label{IC-coupling-fit}
    \begin{tabular}{l|cc||l|cc}
        \hline\hline
        \multirow{2}{*}{CF}                    &\multirow{2}{*}{Experiments}                  &Fitted values                                &\multirow{2}{*}{SCS}                     &\multirow{2}{*}{Experiments}                                                   &Fitted values\\
                                               &                                              &($\alpha=1.4\pm 0.14$)                       &                                         &                                                                               &($\alpha=1.4\pm 0.14$)\\
        \hline
 $K^{*-}\rho^+$                         &$(6.5\pm 2.5)\%$\cite{ARGUS:1992gpk}          &$(6.58_{-0.10}^{+0.14})\%$                     &$K^{*0}\bar{K}^{*0}[S]$                  &$(5.04\pm 0.30)\times10^{-4}$\cite{dArgent:2017gzv}                            &$(5.31_{-2.04}^{+3.02})\times 10^{{-4}}$\\
        \hline 
        \multirow{2}{*}{$\bar{K}^{*0}\rho^0$}  &$(1.515\pm 0.075)\%$\cite{BESIII:2017jyh}     &\multirow{2}{*}{$(1.53_{-0.26}^{+0.24})\%$}    &\multirow{2}{*}{$K^{*0}\bar{K}^{*0}[P]$} &\multirow{2}{*}{$(2.70\pm 0.18)\times 10^{-4}$\cite{dArgent:2017gzv}}          &\multirow{2}{*}{$(2.82_{-0.13}^{+0.08})\times 10^{-4}$}\\
                                               &$(1.59\pm 0.35)\%$\cite{MARK-III:1991fvi}     &&&\\
        \hline
 $\bar{K}^{*0}\rho^0[T]$                &$(1.8\pm 0.6)\%$\cite{MARK-III:1991fvi}       &$(1.10_{-0.16}^{+0.13})\%$                     &$K^{*0}\bar{K}^{*0}[D]$                  &$(1.06\pm 0.09)\times 10^{-4}$\cite{dArgent:2017gzv}                           &$(0.11_{-0.03}^{+0.04})\times 10^{-4}$\\
        \hline
 $\bar{K}^{*0}\omega$                   &$(1.1\pm 0.5)\%$\cite{ARGUS:1992gpk}          &$(0.95_{-0.06}^{+0.04})\%$                     &$K^{*0}\bar{K}^{*0}[\text{Total}]$       &$(8.80\pm 0.36)\times 10^{-4}$\cite{dArgent:2017gzv}                           &$(8.31_{-2.24}^{+3.18})\times 10^{-4}$\\
        \hline      
    \end{tabular}
\end{table}

\subsection{Polarization and partial-wave b.r.s}

In Tab.~\ref{tab: CF&SCS BR} we present our model calculations of all the CF and SCS channels {\it with} and {\it without} the FSIs in comparison with the experimental data and other theoretical calculations which can help to clarify the role played by the long-distance mechanism. 

Note that it is insufficient for disentangling the role played by the long-distance mechanism given that only the $D^0\to \bar{K}^*V$ ($V=\rho^+,\ \rho^0,\ \omega$) channels are considered. The latter two channels can be connected by the isospin relation and the interferences between the CS and IC can account for their difference by adjusting the IC coupling parameter. However, the combined analysis can give clear evidences for the FSIs and we highlight some of the key observations below:

(I) For $D^0\to \phi\rho^0$ and $\phi\omega$, where the IC transition does not contribute, the FSIs can naturally explain the ordering of their total b.r.s and provide a cancellation mechanism for the longitudinal polarization in the $\phi\omega$ channel.

\begin{figure}[H]
    \centering
        \includegraphics[width=8cm]{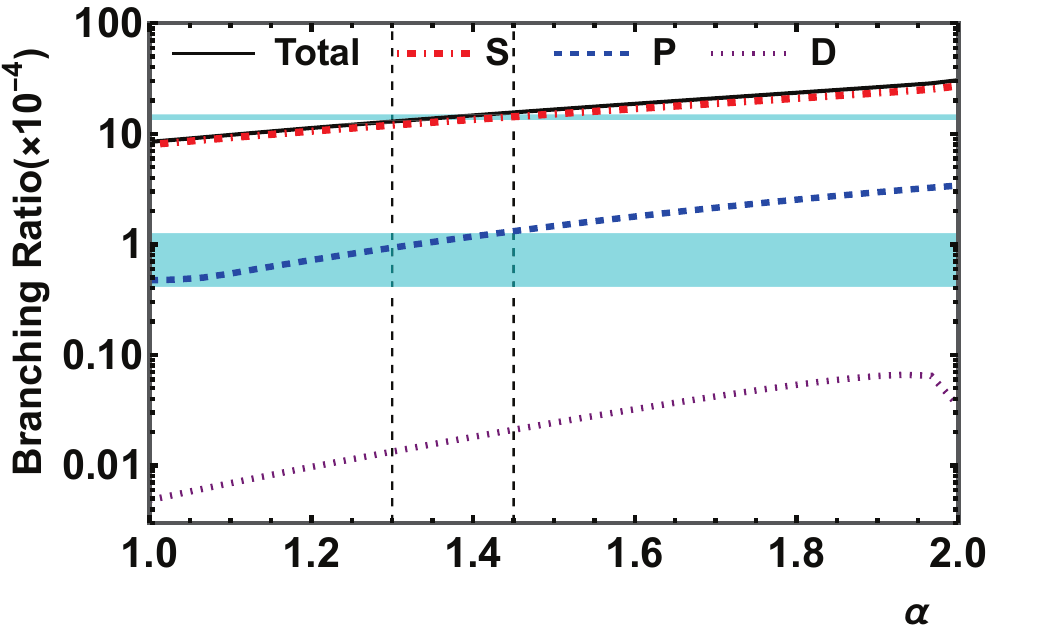}
        \includegraphics[width=8cm]{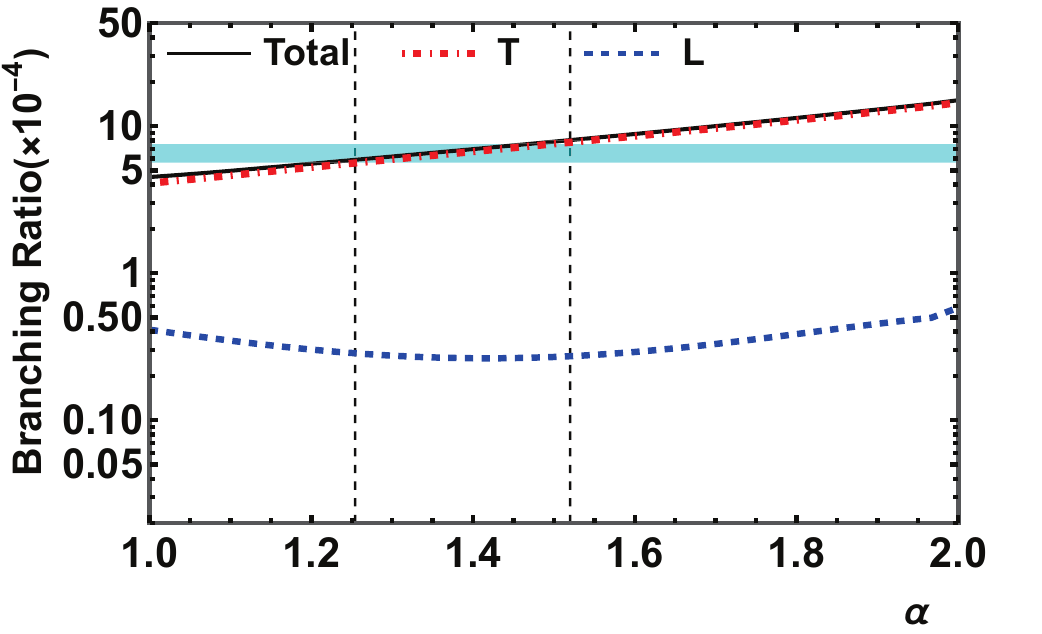}
    \caption{Cut-off parameter $\alpha$ dependence of (a) the partial-wave b.r.s of $D^0\to \phi\rho^0$, and (b) polarization b.r.s of $\phi\omega$, respectively. The solid lines stand for the total b.r.s. The partial-wave and polarization b.r.s are denoted by the line legends.}
    \label{fig:phi-V}
  \end{figure}

In Fig.~\ref{fig:phi-V} we plot the cut-off parameter $\alpha$ dependence of the partial-wave b.r.s of $D^0\to \phi\rho^0$ in (a) and polarization b.r.s of $\phi\omega$ in (b), respectively, to compare with the experimental data (horizontal bands)~\cite{dArgent:2017gzv,BESIII:2021raf}. The vertical lines indicate the range of $\alpha$ with which the experimental data can be well reproduced. 

In Fig.~\ref{fig:phi-V} (a) the dot-dashed ($S$-wave), dashed ($P$-wave) and dotted line ($D$-wave) denote the cut-off parameter $\alpha$ dependence of the calculated partial-wave b.r.s of $D^0\to \phi\rho^0$ in comparison with the experimental data~\cite{dArgent:2017gzv} (horizontal bands). The dominance of the $S$-wave is confirmed which means that the transition is dominantly via the PV processes. Note that the data have quite large errors and the $P$-wave ($(0.08\pm 0.04)\times 10^{-4}$) and $D$-wave ($(0.08\pm 0.03)\times 10^{-4}$) bands almost exactly overlap with each other. Due to the large errors with the $P$ and $D$-wave data, more precise measurements are needed in the future, though it does not affect the main conclusion of the $S$-wave dominance. We also calculate the polarization b.r.s for $D^0\to \phi\rho^0$ and the results are presented in Tab.~\ref{tab: CF&SCS BR}. It shows that the transverse polarization b.r. is about 2 times larger than the longitudinal one. This feature is different from the observations of $D^0\to \phi\omega$, where the longitudinal b.r. turns out to be much smaller than the transverse one~\cite{BESIII:2021raf}.

In Fig.~\ref{fig:phi-V} (b) the cut-off dependence of the $D^0\to \phi\omega$ polarization b.r.s are shown by the dot-dashed (transverse) and dashed line (longitudinal) in comparison with the data (horizontal band). Note that our result for the longitudinal polarization b.r.  $(0.03^{+0.001}_{-0.002})\times 10^{-3}$ is small enough to be accommodated by the data. In terms of the longitudinal polarization fraction $f_L$, we have $f_L\simeq 0.045^{+0.011}_{-0.009}$, which is consistent with the BESIII measurement, i.e. $f_L<0.24$ at 95\% confidence level~\cite{BESIII:2021raf}. We note that all the other existing calculations have predicted comparable (or equal) b.r.s for $D^0\to \phi\rho^0$ and $\phi\omega$~\cite{Cheng:2010rv,Jiang:2017zwr,Uppal:1992se,Bauer:1986bm,Kamal:1990ky,Hinchliffe:1995hz,Bedaque:1993fb}.

(II) A combined view of $D^0\to \rho^0\rho^0$, $\omega\omega$, and $\rho^0\omega$ can be gained. As shown by Tab.~\ref{table:short-dist-amp}, the CS and IC amplitudes have a constructive phase in the $\rho^0\rho^0$ channel, but destructive in $\omega\omega$. It thus predicts a small b.r. for  $\omega\omega$. Significant enhancement comes from the $K^{*+}K^{*-}$ and $\rho^+\rho^-$ rescatterings, and the numerical results in Tab.~\ref{tab: CF&SCS BR} for $\omega\omega$ provides a quantitative estimate of the FSI effects in this channel. An interesting feature with the $\rho^0\omega$ channel is that its CS amplitudes actually cancel out and only the IC amplitude can contribute as the leading short-distance mechanism. However, due to the destructive interference from the FSIs its total b.r. is predicted to be $\sim (0.21^{+0.009}_{-0.005})\times 10^{-3}$ which is comparable with $\rho^0\rho^0$. In comparison with other model calculations our results for $\rho^+\rho^-$, $\rho^0\rho^0$ and $\omega\omega$ are in agreement with Ref.~\cite{Cheng:2010rv}, but quite different from other calculations~\cite{Jiang:2017zwr,Uppal:1992se,Bauer:1986bm,Kamal:1990ky,Hinchliffe:1995hz,Bedaque:1993fb}. Note that Ref.~\cite{Cheng:2010rv} does not calculate the $\rho^0\omega$ channel, while our prediction for $\rho^0\omega$ is very different from other existing models. Hence, a systematic measurement of these non-strange $VV$ channels at BESIII can provide a test of our model. 

(III) Another interesting observation about $D^0\to VV$ is that the SCS decays of $K^{*+}K^{*-}$ and $\rho^+\rho^-$ have not been measured in experiment so far. Since they involve the DE transitions, their decay b.r.s are expected to be sizeable and they should be among the most important decay channels for $D^0$. Also, the polarization and/or partial-wave b.r.s of the CF channel ${K}^{*-}\rho^+$ are unavailable. Although theoretical estimates can be found in the literature, experimental data will provide a better constraint on the NRCQM input in our model. These channels can be accessed by the BESIII experiment and analyses of these channels are strongly recommended.

\begin{table}[H]
    \centering
    \caption{The calculated polarization and partial-wave b.r.s of all the CF and SCS decays of $D^0\to VV$ in units of $10^{-3}$. Columns 3-10 are the results of other theoretical models, including factorization approach~\cite{Cheng:2010rv,Jiang:2017zwr,Bauer:1986bm,Uppal:1992se,Kamal:1990ky}, flavor SU(3) symmetry model (asterisk)~\cite{Kamal:1990ky}, broken flavor SU(3) symmetry model~\cite{Hinchliffe:1995hz} and pole-dominance model~\cite{Bedaque:1993fb}, and the values in parentheses are the results the FSIs considered. The last column lists the experimental data, while the last second and third columns list our model calculations {\it without} and {\it with} the FSIs, respectively.}\label{tab: CF&SCS BR}
    \scalebox{0.92}{
    \begin{tabular}{ccccccccccccc}
    \hline\hline
    \multirow{2}{*}{Process}             &                       &\multirow{2}{*}{\cite{Jiang:2017zwr}}&\multirow{2}{*}{\cite{Cheng:2010rv}}&\multirow{2}{*}{\cite{Uppal:1992se}}&\multirow{2}{*}{\cite{Bauer:1986bm}}&\multirow{2}{*}{\cite{Kamal:1990ky}}&\multirow{2}{*}{\cite{Kamal:1990ky}$^*$}&\multirow{2}{*}{\cite{Hinchliffe:1995hz}}&\multirow{2}{*}{\cite{Bedaque:1993fb}}&{Our results}&Our results&\multirow{2}{*}{Experiments}      \\
    &&&&&&&&&&{\it {without}} FSIs&{\it {with}} FSIs\\
    \hline
    \multirow{3}{*}{$K^{*-}\rho^+$}      &T                      &-                            &-                            &-                               &-                          &-                             &-                             &-                                     &-                                   &$47.59$                        &$57.92_{-0.59}^{+0.96}$                  &-\\
                                         &L                      &$34.7\pm 1.4$                &-                            &-                               &-                          &-                             &-                             &-                                     &-                                   &$4.58$                         &$7.88_{-0.43}^{+0.43}$                   &-\\
                                         &Total                  &-                            &$113$                        &$8.0(77.2)$                     &$236$                      &$65.5(55.9)$                  &-                             &$59\pm 24$                            &$65$                                &$52.17$                        &$65.80_{-1.02}^{+1.39}$                  &$65.0\pm 25.0$~\cite{ARGUS:1992gpk}\\
   \hline
   \multirow{3}{*}{$\bar{K}^{*0}\rho^0$} &T                      &-                            &-                            &-                               &-                          &-                             &-                             &-                                     &-                                   &$12.36$                        &$10.95_{-1.55}^{+1.28}$                  &$18.0\pm 6.0$~\cite{MARK-III:1991fvi}\\
                                         &L                      &$13.2\pm 1.3$                &-                            &-                               &-                          &-                             &-                             &-                                     &-                                   &$5.31$                         &$4.34_{-1.09}^{+1.09}$                   &-\\
                                         &\multirow{2}{*}{Total} &\multirow{2}{*}{-}           &\multirow{2}{*}{$18$}        &\multirow{2}{*}{$8.2(26.0)$}    &\multirow{2}{*}{$22.9$}    &\multirow{2}{*}{$7.0(16.6)$}  &-                             &\multirow{2}{*}{$16\pm 4$}            &\multirow{2}{*}{$8.5$}              &\multirow{2}{*}{$17.68$}       &\multirow{2}{*}{$15.29_{-2.64}^{+2.37}$} &$15.9\pm 3.5$~\cite{MARK-III:1991fvi}\\
                                         &                       &                             &                             &                                &                           &                              &                              &                                      &                                    &                                                 &                                &$15.15\pm 0.75$~\cite{BESIII:2017jyh} \\
   \hline
   \multirow{3}{*}{$\bar{K}^{*0}\omega$} &T                      &-                            &-                            &-                               &-                          &-                             &-                             &-                                     &-                                   &$7.52$                         &$6.85_{-0.51}^{+0.36}$                   &-\\
                                         &L                      &$34.9\pm 2.7$                &-                            &-                               &-                          &-                             &-                             &-                                     &-                                   &$2.76$                         &$2.62_{-0.08}^{+0.09}$                   &-\\
                                         &Total                  &-                            &$16$                         &$10.0(12.6)$                    &$21.9$                     &$6.6(6.6)$                    &$28\pm 17$                    &$11\pm 5$                             &$7.9$                               &$10.28$                        &$9.48_{-0.59}^{+0.45}$                   &$11.0\pm 5.0$~\cite{ARGUS:1992gpk}\\
   \hline\hline
   \multirow{3}{*}{$K^{*+}K^{*-}$}       &T                      &-                            &-                            &-                               &-                          &-                             &-                             &-                                     &-                                   &$4.02$                         &$6.75_{-0.38}^{+0.26}$                   &-\\
                                         &L                      &$1.1\pm 0.05$                &-                            &-                               &-                          &-                             &-                             &-                                     &-                                   &$1.83$                         &$3.17_{-0.17}^{+0.09}$                   &-\\
                                         &Total                  &-                            &$7.3$                        &$10.0(3.3)$                     &$10.1$                     &$2.4(1.8)$                    &$1.5\pm 0.8$                  &$2.4_{-2.1}^{+4.1}$                   &-                                   &$5.86$                         &$9.92_{-0.55}^{+0.34}$                   &-\\
   \hline 
   \multirow{6}{*}{$K^{*0}\bar{K}^{*0}$} &$\bf{S}$               &-                            &-                            &-                               &-                          &-                             &-                             &-                                     &-                                   &$0.92$                         &$0.53_{-0.20}^{+0.30}$                   &$0.50\pm 0.03$~\cite{dArgent:2017gzv}\\
                                         &$\bf{P}$               &-                            &-                            &-                               &-                          &-                             &-                             &-                                     &-                                   &$0.30$                         &$0.28_{-0.012}^{+0.008}$                 &$0.27\pm 0.02$~\cite{dArgent:2017gzv}\\
                                         &$\bf{D}$               &-                            &-                            &-                               &-                          &-                             &-                             &-                                     &-                                   &$0.006$                        &$0.01_{-0.003}^{+0.004}$                 &$0.11\pm 0.01$~\cite{dArgent:2017gzv}\\
                                         &T                      &-                            &-                            &-                               &-                          &-                             &-                             &-                                     &-                                   &$0.84$                         &$0.58_{-0.13}^{+0.19}$                   &-\\
                                         &L                      &$0.01\pm 0.002$              &-                            &-                               &-                          &-                             &-                             &-                                     &-                                   &$0.39$                         &$0.25_{-0.08}^{+0.11}$                   &-\\
                                         &Total                  &-                            &-                            &$10.0(1.1)$                     &-                          &$0(0.6)$                      &$0.65\pm 0.3$                 &$2.0\pm 1.5$                          &$0.026$                             &$1.23$                         &$0.83_{-0.22}^{+0.31}$                   &$0.88\pm0.04$~\cite{dArgent:2017gzv}\\
   \hline
   \multirow{3}{*}{$\rho^+\rho^-$}       &T                      &-                            &-                            &-                               &-                          &-                             &-                             &-                                     &-                                   &$5.44$                         &$5.80_{-0.36}^{+0.36}$                   &-\\
                                         &L                      &$3.2\pm 0.1$                 &-                            &-                               &-                          &-                             &-                             &-                                     &-                                   &$1.36$                         &$3.22_{-0.03}^{+0.004}$                  &- \\
                                         &Total                  &-                            &$6.6$                        &$7.3(6.2)$                      &$13.1$                     &$5.3(4.4)$                    &$5.4\pm 3.2$                  &$<15$                                 &-                                   &$6.81$                         &$9.03_{-0.34}^{+0.36}$                   &-\\
   \hline
   \multirow{7}{*}{$\rho^0\rho^0$}       &$\bf{S}$               &-                            &$0.85$                       &-                               &-                          &-                             &-                             &-                                     &-                                   &$0.49$                         &$0.45_{-0.26}^{+0.40}$                   &$0.18\pm 0.13$~\cite{dArgent:2017gzv}\\
                                         &$\bf{P}$               &-                            &$0.091$                      &-                               &-                          &-                             &-                             &-                                     &-                                   &$0.23$                         &$0.56_{-0.10}^{+0.10}$                   &$0.53\pm 0.13$~\cite{dArgent:2017gzv}\\
                                         &$\bf{D}$               &-                            &$0.034$                      &-                               &-                          &-                             &-                             &-                                     &-                                   &$0.01$                         &$0.03_{-0.01}^{+0.01}$                   &$0.62\pm 0.30$~\cite{dArgent:2017gzv}\\
                                         &T                      &-                            &-                            &-                               &-                          &-                             &-                             &-                                     &-                                   &$0.48$                         &$0.87_{-0.25}^{+0.32}$                   &$0.56\pm 0.07$~\cite{FOCUS:2007ern}\\
                                         &L                      &$1.1\pm 0.1$                 &-                            &-                               &-                          &-                             &-                             &-                                     &-                                   &$0.25$                         &$0.18_{-0.13}^{+0.19}$                   &$1.27\pm 0.10$~\cite{FOCUS:2007ern}\\
                                         &\multirow{2}{*}{Total} &\multirow{2}{*}{-}           &\multirow{2}{*}{$0.97$}      &\multirow{2}{*}{$7.3(1.6)$}     &\multirow{2}{*}{1.18}      &\multirow{2}{*}{$0.5(1.3)$}   &\multirow{2}{*}{$1.7\pm 1.0$} &\multirow{2}{*}{$<6.5$}               &\multirow{2}{*}{-}                  &\multirow{2}{*}{0.73}          &\multirow{2}{*}{$1.05_{-0.37}^{+0.50}$}  &$1.85\pm 0.13$~\cite{FOCUS:2007ern}\\
                                         &                       &                             &                             &                                &                           &                              &                              &                                      &                                    &                               &                                         &$1.33\pm 0.35$~\cite{dArgent:2017gzv}\\
   \hline
   \multirow{3}{*}{$\omega\omega$}       &T                      &-                            &-                            &-                               &-                          &-                             &-                             &-                                     &-                                   &$0.019$                         &$0.12_{-0.017}^{+0.018}$                &-\\
                                         &L                      &$0.47\pm 0.07$               &-                            &-                               &-                          &-                             &-                             &-                                     &-                                   &$0.00065$                       &$0.03_{-0.001}^{+0.0005}$               &-\\
                                         &Total                  &-                            &$0.68$                       &-                               &$1.08$                     &$0.2(0.2)$                    &$2.3\pm 1.4$                  &-                                     &-                                   &$0.020$                         &$0.15_{-0.018}^{+0.018}$                &-\\
   \hline
   \multirow{3}{*}{$\rho^0\omega$}       &T                      &-                            &-                            &-                               &-                          &-                             &-                             &-                                     &-                                   &$0.84$                          &$0.15_{-0.002}^{+0.013}$                &-\\
                                         &L                      &$0.95\pm 0.07$               &-                            &-                               &-                          &-                             &-                             &-                                     &-                                   &$0.13$                          &$0.06_{-0.008}^{+0.004}$                &-\\
                                         &Total                  &-                            &-                            &$0.03$                          &-                          &$0.02(0.02)$                  &$3.0\pm 1.8$                  &$<84$                                 &-                                   &$0.97$                          &$0.21_{-0.005}^{+0.009}$                &-\\
   \hline
   \multirow{6}{*}{$\phi\rho^0$}         &$\bf{S}$               &-                            &$0.63$                       &-                               &-                          &-                             &-                             &-                                     &-                                   &$0.48$                          &$1.35_{-0.20}^{+0.23}$                  &$1.40\pm 0.12$~\cite{dArgent:2017gzv}\\
                                         &$\bf{P}$               &-                            &$0.025$                      &-                               &-                          &-                             &-                             &-                                     &-                                   &$0.05$                          &$0.11_{-0.03}^{+0.04}$                  &$0.08\pm 0.04$~\cite{dArgent:2017gzv}\\
                                         &$\bf{D}$               &-                            &$0.001$                      &-                               &-                          &-                             &-                             &-                                     &-                                   &$\sim 0$                        &$0.002_{-0.001}^{+0.001}$               &$0.08\pm 0.03$~\cite{dArgent:2017gzv}\\
                                         &T                      &-                            &-                            &-                               &-                          &-                             &-                             &-                                     &-                                   &$0.37$                          &$1.02_{-0.18}^{+0.21}$                  &-  \\
                                         &L                      &$0.65\pm 0.04$               &-                            &-                               &-                          &-                             &-                             &-                                     &-                                   &$0.16$                          &$0.45_{-0.06}^{+0.07}$                  &-  \\
                                         &Total                  &-                            &$0.66$                       &$7.6(0.4)$                      &$1.02$                     &$0.26(0.26)$                  &$0.038\pm 0.014$              &$1.9\pm 0.5$                          &$0.22$                              &$0.53$                          &$1.47_{-0.24}^{+0.27}$                  &$1.56\pm0.13$~\cite{dArgent:2017gzv}\\
    \hline
    \multirow{3}{*}{$\phi\omega$}        &T                      &-                            &-                            &-                               &-                          &-                             &-                             &-                                     &-                                   &$0.34$                          &$0.67_{-0.10}^{+0.12}$                  &$0.65\pm 0.10$~\cite{BESIII:2021raf}\\
                                         &L                      &$1.41\pm 0.09$               &-                            &-                               &-                          &-                             &-                             &-                                     &-                                   &$0.15$                          &$0.03_{-0.002}^{+0.001}$                &$\sim 0$~\cite{BESIII:2021raf}\\
                                         &Total                  &-                            &$0.66$                       &-                               &$0.92$                     &$0.23(0.23)$                  &$0.035\pm 0.13$               &-                                     &-                                   &$0.49$                          &$0.69_{-0.10}^{+0.12}$                  &$0.65\pm 0.10$~\cite{BESIII:2021raf}\\
    \hline
    \end{tabular}
    }
\end{table}

\section{Summary} 
In this work we carry out a systematic analysis of the CF and SCS decays of $D^0\to VV$ by taking into account the long-distance FSIs as a crucial mechanism for understanding the mysterious polarization puzzles. We show that the NRCQM provides a reasonably well description of the DE and CS transitions with explicit phase constraints. The IC transition contains more profound effects arising from the complicated intermediate configurations. In our approach it can be well parametrized out with the FSIs considered, and can be determined by the experimental data. Our analysis shows that the stunning discrepancies of the decay rates between $D^0\to \phi\rho^0$ and $\phi\omega$, and the unexpectedly small longitudinal polarization b.r. of the $\phi\omega$ channel can be naturally explained by the FSIs. It provides a clear evidence for such a long-distance mechanism in $D$ meson decays. We also strongly recommend future precise and completed measurements of $D^0\to VV$ at BESIII since it will provide us a unique probe for resolving some of those profound non-perturbative dynamics.

\section*{Acknowledgement}

Authors would like to thanks Prof. Hai-Yang Cheng for useful discussions. This work is supported, in part, by the National Natural Science Foundation of China (Grant No. 12235018),  DFG and NSFC funds to the Sino-German CRC 110 ``Symmetries and the Emergence of Structure in QCD'' (NSFC Grant No. 12070131001, DFG Project-ID 196253076), National Key Basic Research Program of China under Contract No. 2020YFA0406300, and Strategic Priority Research Program of Chinese Academy of Sciences (Grant No. XDB34030302).

\appendix

\section{Transition amplitudes }\label{append}

There are three nonvanishing helicity amplitudes for $D^0\to VV$: $\mathcal{M}^{++}$, $\mathcal{M}^{--}$ and $\mathcal{M}^{00}$, where the superscripts ``$\pm$" and ``$0$" denote the helicity of the final vector meson along the momentum direction of one of the final vector meson. The nonvanishing amplitudes, $\mathcal{M}^{++}$, $\mathcal{M}^{--}$, are not independent, and symmetry connects them via $\mathcal{M}^{++}_{\text{PC}}=-\mathcal{M}^{--}_{\text{PC}}$ and $\mathcal{M}^{++}_{\text{PV}}=\mathcal{M}^{--}_{\text{PV}}$. Other amplitudes $\mathcal{M}^{\pm\mp}$ and $\mathcal{M}_{PC}^{00}$ vanish. 

\subsection{Short-distance transition amplitudes extracted in the quark model}\label{appendix:1}

The transition amplitudes of the DE and CS processes as the short-distance dynamics are calculated in the NRCQM~\cite{Kokoski:1985is,Godfrey:1985xj,Godfrey:1986wj} and the operators have been extracted in Refs.~\cite{LeYaouanc:1988fx, Richard:2016hac, Niu:2020gjw}. 
The transition amplitudes for the DE and CS processes are listed below for different channels:
    \begin{itemize}
        \item DE-process
        \begin{align}
            \mathcal{M}^{++}_{\text{PC}}(D^0\to K^{*-}\rho^+)&=\frac{G_FV_{cs}V_{ud}R_D^{3/2}R_K^{3/2}R_{\rho}^{3/2}\left((m_c+m_q)m_sR_D^2+(m_s+m_q)m_cR_K^2\right)p}{2\pi^{9/4}m_cm_s(m_s+m_q)(R_D^2+R_K^2)^{5/2}}\\
            &\times\text{exp}\left(-\frac{m_q^2p^2}{2(m_s+m_q)^2(R_D^2+R_K^2)}\right),\\
            \mathcal{M}_{\text{PV}}^{++}(D^0\to K^{*-}\rho^+)&=\frac{G_FV_{cs}V_{ud}R_D^{3/2}R_K^{3/2}R_{\rho}^{3/2}}{\pi^{9/4}(R_D^2+R_K^2)^{3/2}}\text{exp}\left(-\frac{m_q^2p^2}{2(m_s+m_q)^2(R_D^2+R_K^2)}\right),\\
            \mathcal{M}_{\text{PV}}^{00}(D^0\to K^{*-}\rho^+)&=-\frac{G_FV_{cs}V_{ud}R_D^{3/2}R_K^{3/2}R_{\rho}^{3/2}}{\pi^{9/4}(R_D^2+R_K^2)^{3/2}}\text{exp}\left(-\frac{m_q^2p^2}{2(m_s+m_q)^2(R_D^2+R_K^2)}\right),\\
            \mathcal{M}(D^0\to K^{*+}K^{*-})&=\mathcal{M}(D^0\to K^{*-}\rho^+)[R_{\rho}\overset{\text{replace}}{\to} R_K,\ V_{ud}\overset{\text{replace}}{\to}V_{us}]\\
            \mathcal{M}(D^0\to \rho^+\rho^-)&=\mathcal{M}(D^0\to K^{*-}\rho^+)[R_{K}\overset{\text{replace}}{\to} R_\rho,\ m_s\overset{\text{replace}}{\to}m_q,\ V_{cs}\overset{\text{replace}}{\to}V_{cd}].
        \end{align}
        \item CS-process
        \begin{align}
            \mathcal{M}_{\text{PC}}^{++}(D^0\to\bar{K^{*0}}\rho^0)&=\frac{G_FV_{cs}V_{ud}(R_DR_{K}R_{\rho})^{3/2}\left((m_c+m_q)R_D^2+2m_cR_{\rho}^2\right)p}{12\sqrt{2}\pi^{9/4}m_cm_q(R_D^2+R_{\rho}^2)^{5/2}}\text{exp}\left(-\frac{p^2}{8(R_D^2+R_{\rho}^2)}\right),\\
            \mathcal{M}_{\text{PV}}^{++}(D^0\to\bar{K^{*0}}\rho^0)&=\frac{G_FV_{cs}V_{ud}(R_DR_{K}R_{\rho})^{3/2}}{3\sqrt{2}\pi^{9/4}(R_D^2+R_{\rho}^2)^{3/2}}\text{exp}\left(-\frac{p^2}{8(R_D^2+R_{\omega/\rho}^2)}\right),\\
            \mathcal{M}_{\text{PV}}^{00}(D^0\to\bar{K^{*0}}\rho^0)&=-\frac{G_FV_{cs}V_{ud}(R_DR_{K}R_{\rho})^{3/2}}{3\sqrt{2}\pi^{9/4}(R_D^2+R_{\rho}^2)^{3/2}}\text{exp}\left(-\frac{p^2}{8(R_D^2+R_{\omega/\rho}^2)}\right),\\
            \mathcal{M}(D^0\to\bar{K^{*0}}\omega)&=\mathcal{M}(D^0\to\bar{K^{*0}}\rho^0)[R_{\rho}\overset{\text{replace}}{\to}R_{\omega}],\\
            \mathcal{M}(D^0\to\phi\rho^0)&=\mathcal{M}(D^0\to\bar{K^{*0}}\rho^0)[R_K\overset{\text{replace}}{\to}R_{\phi},\ V_{ud}\overset{\text{replace}}{\to}V_{us}],\\
            \mathcal{M}(D^0\to\phi\omega)&=\mathcal{M}(D^0\to\bar{K^{*0}}\omega)[R_K\overset{\text{replace}}{\to}R_{\omega},\ V_{ud}\overset{\text{replace}}{\to}V_{us}],\\
            \mathcal{M}(D^0\to\rho^0\rho^0)&=-\frac{1}{\sqrt{2}}\mathcal{M}(D^0\to\bar{K}^{*0}\rho^0)[R_K\overset{\text{replace}}{\to}R_{\rho},\ V_{cs}\overset{\text{replace}}{\to}V_{cd}],\\
            \mathcal{M}(D^0\to\omega\omega)&=\frac{1}{\sqrt{2}}\mathcal{M}(D^0\to\bar{K}^{*0}\omega)[R_K\overset{\text{replace}}{\to}R_{omega},\ V_{cs}\overset{\text{replace}}{\to}V_{cd}].
        \end{align}

    \end{itemize}
In the above equations, ${p}\equiv|\boldsymbol{p}|$ denotes the three-vector momentum of the final vector mesons in the initial-state rest frame; $m_q$ is the mass of the light quarks $(u, d)$, $m_s$ and $m_c$ represent the masses of the $s$ and $c$ quark, respectively; $R_D$, $R_K$, $R_{\omega/\rho}$ and $R_{\phi}$ are the harmonic oscillator (H.O.) strengths determined by the ground-state mesons $D^0$, $K^*$, $\omega/\rho^0$ and $\phi$, respectively. The values adopted for these quark model parameters are from Refs.~\cite{Kokoski:1985is,Godfrey:1985xj,Godfrey:1986wj} and they are listed in Tab.~\ref{QM-para}.

\begin{table}[H]
    \centering
    \caption{Input values of the constituent quark masses and harmonic oscillator strengths adopted in our calculations, which are from Refs.~\cite{Kokoski:1985is,Godfrey:1985xj,Godfrey:1986wj}.}\label{QM-para}
    \begin{tabular}{c|c||c|c}
        \hline\hline
        H.O. strength    &Values [GeV]    &Quark mass    &Values [GeV]\\
        \hline
        $R_D$                  &$0.66$          &$m_c$         &$1.628$ \\
        \hline
        $R_{K^*}$              &$0.48$          &$m_s$         &$0.419$ \\
        \hline
        $R_{\rho/\omega}$      &$0.45$          &$m_q$         &$0.22$ \\
        \hline
        $R_{\phi}$             &$0.51$          &-&- \\
        \hline 
    \end{tabular}
\end{table}


\subsection{Long-distance transition amplitudes from FSIs}\label{appendix:2}

In this section, we present the loop amplitudes for the convenience of tracking the calculation details. For
simplicity, we do not distinguish the coupling constants at the hadronic vertices but just denote them as $g_i$ with $i=1, 2, 3$. The amplitudes for different processes are listed explicitly as follows, and we employ {\it LoopTools} (\url{https://www.feynarts.de/looptools/}) to accomplish the numerical calculations:
    \begin{itemize}
        \item $\tilde{\mathcal{I}}[(\text{PC}),\ K^*,\ \bar{K}^*,\ (K)]$
            \begin{align}
                \begin{split}
            i\mathcal{M}=&g_1g_2g_3\int\frac{d^4p_1}{(2\pi)^4}\frac{\epsilon_{\alpha\beta\mu\nu}p_1^{\alpha}p_3^{\beta}(g^{\mu\mu^{\prime}}-\frac{p_1^{\mu}p_1^{\mu^{\prime}}}{p_1^2})\epsilon_{\alpha_1\beta_1\mu^{\prime}\delta}p_1^{\alpha_1}p_{V_1}^{\beta_1}\varepsilon_{V_1}^{\delta*}\epsilon_{\alpha_2\beta_2\nu^{\prime}\lambda}p_3^{\alpha_2}p_{V_2}^{\beta_2}\varepsilon_{V_2}^{\lambda*}(g^{\nu\nu^{\prime}}-\frac{p_3^{\nu}p_3^{\nu^{\prime}}}{p_3^2})}{(p_1^2-m_{K^*}^2+i\epsilon)(p_2^2-m_K^2+i\epsilon)(p_3^2-m_{K^*}^2+i\epsilon)}\mathcal{F}(p_i^2)\\
            &=g_1g_2g_3\int\frac{d^4p_1}{(2\pi)^4}\frac{\epsilon_{\alpha\beta\mu\nu}p_1^{\alpha}p_3^{\beta}\times\epsilon_{\alpha_1\beta_1\mu\delta}p_1^{\alpha_1}p_{V_1}^{\beta_1}\varepsilon_{V_1}^{\delta*}\times\epsilon_{\alpha_2\beta_2\nu\lambda}p_3^{\alpha_2}p_{V_2}^{\beta_2}\varepsilon_{V_2}^{\lambda*}}{(p_1^2-m_{K^*}^2+i\epsilon)(p_2^2-m_K^2+i\epsilon)(p_3^2-m_{K^*}^2+i\epsilon)}\mathcal{F}(p_i^2)\\
            &=g_1g_2g_3\int\frac{d^4p_1}{(2\pi)^4}\frac{\mathcal{F}(p_i^2)}{(p_1^2-m_{K^*}^2+i\epsilon)(p_2^2-m_K^2+i\epsilon)(p_3^2-m_{K^*}^2+i\epsilon)}\\
            &\times\{\epsilon_{\alpha\beta\delta\lambda}p_{V_1}^{\alpha}p_{V_2}^{\beta}\varepsilon_{V_1}^{\delta*}\varepsilon_{V_2}^{\lambda*}[(p_1\cdot p_{V_1})^2+(p_1\cdot p_{V_1})(p_1\cdot p_{V_2})-p_1^2(p_{V_1}^2+p_{V_1}\cdot p_{V_2})]\\
            &+\epsilon_{\alpha\beta\delta\lambda}p_1^{\alpha}p_{V_2}^{\beta}\varepsilon_{V_1}^{\delta*}\varepsilon_{V_2}^{\lambda*}[p_1^2p_{V_1}^2-(p_1\cdot p_{V_1})^2]\\
            &+\epsilon_{\alpha\beta\delta\lambda}p_{V_1}^{\alpha}p_1^{\beta}\varepsilon_{V_1}^{\delta*}\varepsilon_{V_2}^{\lambda*}[-p_1^2p_{V_2}^2+(p_1\cdot p_{V_2})^2]\\
            &+\epsilon_{\alpha\beta\delta\lambda}p_{V_1}^{\alpha}p_{V_2}^{\beta}p_1^{\delta}\varepsilon_{V_2}^{\lambda*}[(p_1\cdot\varepsilon_{V_1}^*)(p_{V_1}\cdot p_{V_2}+p_{V_1}^2)-(p_1\cdot p_{V_1})(p_1\cdot\varepsilon_{V_1}^*+p_{V_2}\cdot \varepsilon_{V_1}^*)]\\
            &+\epsilon_{\alpha\beta\delta\lambda}p_{V_1}^{\alpha}p_{V_2}^{\beta}\varepsilon_{V_1}^{\delta*}p_1^{\lambda}(p_1\cdot p_{V_2})(p_1\cdot\varepsilon_{V_2}^*)\}
                \end{split}
            \end{align}

        \item $\tilde{\mathcal{I}}[\text{(PC)},\ K^*,\ \bar{K}^*,\ (K^*)]$
        \begin{align}
            \begin{split}
                i\mathcal{M}
                &=g_1g_2g_3\int\frac{d^4p_1}{(2\pi)^4}\frac{\epsilon_{\alpha\beta\mu\nu}p_1^{\alpha}p_3^{\beta}(g^{\mu\mu^{\prime}}-\frac{p_1^{\mu}p_1^{\mu^{\prime}}}{p_1^2})(g^{\rho\sigma}-\frac{p_2^{\rho}p_2^{\sigma}}{p_2^2})(g^{\nu\nu^{\prime}}-\frac{p_3^{\nu}p_3^{\nu^{\prime}}}{p_3^2})}{(p_1^2-m_{K^*}^2+i\epsilon)(p_2^2-m_{K^*}^2+i\epsilon)(p_3^2-m_{K^*}^2+i\epsilon)}\\
                &\times[(p_1+p_{V_1})_{\rho}\varepsilon_{V_1}^{\delta*}g_{\mu^{\prime}\delta}+(p_2-p_{V_1})_{\mu^{\prime}}\varepsilon_{V_1}^{\delta*}g_{\delta\rho}-(p_1+p_2)_{\delta}\varepsilon_{V_1}^{\delta*}g_{\mu^{\prime}\rho}]\\
                &\times[(p_3+p_{V_2})_{\sigma}\varepsilon_{V_2}^{\lambda*}g_{\nu^{\prime}\lambda}-(p_2+p_{V_2})_{\nu^{\prime}}\varepsilon_{V_2}^{\lambda*}g_{\lambda\sigma}+(p_2-p_3)_{\lambda}\varepsilon_{V_2}^{\lambda*}g_{\nu^{\prime}\sigma}]\mathcal{F}(p_i^2)\\
                &=g_1g_2g_3\int\frac{d^4p_1}{(2\pi)^4}\frac{\epsilon_{\alpha\beta\mu\nu}p_1^{\alpha}p_3^{\beta}(g^{\rho\sigma}-\frac{p_2^{\rho}p_2^{\sigma}}{p_2^2})}{(p_1^2-m_{K^*}^2+i\epsilon)(p_2^2-m_{K^*}^2+i\epsilon)(p_3^2-m_{K^*}^2+i\epsilon)}\mathcal{F}(p_i^2)\\
                &\times[(p_1+p_{V_1})_{\rho}\varepsilon_{V_1}^{\delta*}g_{\mu\delta}+(p_2-p_{V_1})_{\mu}\varepsilon_{V_1}^{\delta*}g_{\delta\rho}-(p_1+p_2)_{\delta}\varepsilon_{V_1}^{\delta*}g_{\mu\rho}]\\
                &\times[(p_3+p_{V_2})_{\sigma}\varepsilon_{V_2}^{\lambda*}g_{\nu\lambda}-(p_2+p_{V_2})_{\nu}\varepsilon_{V_2}^{\lambda*}g_{\lambda\sigma}+(p_2-p_3)_{\lambda}\varepsilon_{V_2}^{\lambda*}g_{\nu\sigma}]\\
                &=g_1g_2g_3\int\frac{d^4p_1}{(2\pi)^4}\frac{4\mathcal{F}(p_i^2)}{p_2^2(p_1^2-m_{K^*}^2+i\epsilon)(p_2^2-m_{K^*}^2+i\epsilon)(p_3^2-m_{K^*}^2+i\epsilon)}\\
                &\times\{\epsilon_{\alpha\beta\delta\lambda}p_1^{\alpha}p_{V_2}^{\beta}\varepsilon_{V_1}^{\delta*}\varepsilon_{V_2}^{\lambda*}[(p_1^2-p_1\cdot p_{V_1})(p_{V_1}\cdot p_{V_2})+(p_{V_1}^2-p_1\cdot p_{V_1})(p_1\cdot p_{V_2})]\\
                &+\epsilon_{\alpha\beta\delta\lambda}p_{V_1}^{\alpha}p_1^{\beta}\varepsilon_{V_1}^{\delta*}\varepsilon_{V_2}^{\lambda*}[(p_1\cdot p_{V_1}-p_1^2)(p_{V_1}\cdot p_{V_2})+(p_1\cdot p_{V_1}-p_{V_1}^2)(p_1\cdot p_{V_2})]\\
                &+\epsilon_{\alpha\beta\delta\lambda}p_{V_1}^{\alpha}p_{V_2}^{\beta}p_1^{\delta}\varepsilon_{V_2}^{\lambda*}p_2^2[-2(p_1\cdot \varepsilon_{V_1}^*)+(p_{V_2}\cdot\varepsilon_{V_1}^*)]\\
                &+\epsilon_{\alpha\beta\delta\lambda}p_{V_1}^{\alpha}p_{V_2}^{\beta}\varepsilon_{V_1}^{\delta*}p_1^{\lambda}p_2^2[-2(p_1\cdot \varepsilon_{V_2}^*)+(p_{V_1}\cdot\varepsilon_{V_2}^*)]\}
            \end{split}
        \end{align}

        \item $\tilde{\mathcal{I}}[\text{(PC)},\ K^*,\ \bar{K}^*,\ (\kappa)]$
        \begin{align}
            \begin{split}
                i\mathcal{M}
                &=g_1g_2g_3\int\frac{d^4p_1}{(2\pi)^4}\frac{\epsilon_{\alpha\beta\mu\nu}p_1^{\alpha}p_3^{\beta}(g^{\mu\rho}-\frac{p_{1}^\mu p_1^\rho}{p_1^2})(g^{\nu\sigma}-\frac{p_3^{\nu}p_3^{\sigma}}{p_3^2})\varepsilon_{V_1}^*\varepsilon_{V_2\sigma}^*}{(p_1^2-m_{K^*}^2+i\epsilon)(p_2^2-m_{\kappa}^2+i\epsilon)(p_3^2-m_{K^*}^2+i\epsilon)}\mathcal{F}(p_i^2)\\
                &=g_1g_2g_3\int\frac{d^4p_1}{(2\pi)^4}\frac{\epsilon_{\alpha\beta\mu\nu}p_1^{\alpha}p_3^{\beta}\varepsilon_{V_1}^{\mu*}\varepsilon_{V_2}^{\nu*}}{(p_1^2-m_{K^*}^2+i\epsilon)(p_2^2-m_{\kappa}^2+i\epsilon)(p_3^2-m_{K^*}^2+i\epsilon)}\mathcal{F}(p_i^2)\\
                &=g_1g_2g_3\int\frac{d^4p_1}{(2\pi)^4}\frac{\mathcal{F}(p_i^2)}{(p_1^2-m_{K^*}^2+i\epsilon)(p_2^2-m_{\kappa}^2+i\epsilon)(p_3^2-m_{K^*}^2+i\epsilon)}\\
                &\times\{\epsilon_{\alpha\beta\delta\lambda}p_1^{\alpha}p_{V_2}^{\beta}\varepsilon_{V_1}^{\delta*}\varepsilon_{V_2}^{\lambda*}-\epsilon_{\alpha\beta\delta\lambda}p_{V_1}^{\alpha}p_1^{\beta}\varepsilon_{V_1}^{\delta*}\varepsilon_{V_2}^{\lambda*}\}
            \end{split}
        \end{align}

        \item $\tilde{\mathcal{I}}[\text{(PC)},\ K,\ \bar{K}^*,\ (K)]$
        \begin{align}
            \begin{split}
                i\mathcal{M}
                &=g_1g_2g_3\int\frac{d^4p_1}{(2\pi)^4}\frac{(p_D+p_1)_{\mu}(p_1+p_2)_{\rho}\epsilon_{\alpha\beta\nu\sigma}p_3^{\alpha}p_{V_2}^{\beta}\varepsilon_{V_1}^{\rho*}\varepsilon_{V_2}^{\sigma*}(g^{\mu\nu}-\frac{p_3^{\mu}p_3^{\nu}}{p_3^2})}{(p_1^2-m_K^2+i\epsilon)(p_2^2-m_K^2+i\epsilon)(p_3^2-m_{K^*}^2+i\epsilon)}\mathcal{F}(p_i^2)\\
                &=g_1g_2g_3\int\frac{d^4p_1}{(2\pi)^4}\frac{(p_D+p_1)^{\mu}(p_1+p_2)_{\rho}\epsilon_{\alpha\beta\mu\sigma}p_3^{\alpha}p_{V_2}^{\beta}\varepsilon_{V_1}^{\rho*}\varepsilon_{V_2}^{\sigma*}}{(p_1^2-m_K^2+i\epsilon)(p_2^2-m_K^2+i\epsilon)(p_3^2-m_{K^*}^2+i\epsilon)}\mathcal{F}(p_i^2)\\
                &=g_1g_2g_3\int\frac{4\mathcal{F}(p_i^2)}{(p_1^2-m_K^2+i\epsilon)(p_2^2-m_K^2+i\epsilon)(p_3^2-m_{K^*}^2+i\epsilon)}\times \epsilon_{\alpha\beta\delta\lambda}p_{V_1}^{\alpha}p_{V_2}^{\beta}p_1^{\delta}\varepsilon_{V_2}^{\lambda}(p_1\cdot\varepsilon_{V_1}^*)
            \end{split}
        \end{align}
        \item $\tilde{\mathcal{I}}[\text{(PC)},\ K^*,\ \bar{K},\ (K)]$
        \begin{align}
            \begin{split}
                i\mathcal{M}
                &=g_1g_2g_3\int\frac{d^4p_1}{(2\pi)^4}\frac{(p_D+p_3)_{\mu}(p_3-p_2)_{\rho}\epsilon_{\alpha\beta\nu\sigma}p_{1}^{\alpha}p_{V_1}^{\beta}\varepsilon_{V_1}^{\sigma*}\varepsilon_{V_2}^{\rho*}(g^{\mu\nu}-\frac{p_1^{\mu}p_1^{\nu}}{p_1^2})}{(p_1^2-m_{K^*}^2+i\epsilon)(p_2^2-m_K^2+i\epsilon)(p_3^2-m_K^2+i\epsilon)}\mathcal{F}(p_i^2)\\
                &=g_1g_2g_3\int\frac{d^4p_1}{(2\pi)^4}\frac{(p_D+p_3)^{\mu}(p_3-p_2)_{\rho}\epsilon_{\alpha\beta\mu\sigma}p_{1}^{\alpha}p_{V_1}^{\beta}\varepsilon_{V_1}^{\sigma*}\varepsilon_{V_2}^{\rho*}}{(p_1^2-m_{K^*}^2+i\epsilon)(p_2^2-m_K^2+i\epsilon)(p_3^2-m_K^2+i\epsilon)}\mathcal{F}(p_i^2)\\
                &=g_1g_2g_3\int\frac{d^4p_1}{(2\pi)^4 }\frac{4\mathcal{F}(p_i^2)}{(p_1^2-m_{K^*}^2+i\epsilon)(p_2^2-m_K^2+i\epsilon)(p_3^2-m_K^2+i\epsilon)}\times\epsilon_{\alpha\beta\delta\lambda}p_{V_1}^{\alpha}p_{V_2}^{\beta}\varepsilon_{V_1}^{\delta*}p_1^{\lambda}(p_1\cdot\varepsilon_{V_2}^*-p_{V_1}\cdot\varepsilon_{V_2}^*)
            \end{split}
        \end{align}
        \item $\tilde{\mathcal{I}}[\text{(PC)},\ K,\ \bar{K}^*,\ (K^*)]$
        \begin{align}
            \begin{split}
                i\mathcal{M}&=g_1g_2g_3\int\frac{d^4p_1}{(2\pi)^4}\frac{(p_D+p_1)_{\nu}\varepsilon_{\alpha\beta\rho\delta}p_2^{\alpha}p_{V_1}^{\beta}\varepsilon_{V_1}^{\delta*}(g^{\mu\nu}-\frac{p_3^{\mu}p_3^{\nu}}{p_3^2})(g^{\rho\sigma}-\frac{p_2^{\rho}p_2^{\sigma}}{p_2^2})}{(p_1^2-m_K^2+i\epsilon)(p_2^2-m_{K^*}^2+i\epsilon)(p_3^2-m_{K^*}^2+i\epsilon)}\\
                &\times[(p_3+p_{V_2})_{\sigma}\varepsilon_{V_2\mu}^*+(p_2-p_3)_{\lambda}\varepsilon_{V_2}^{\lambda*}g_{\mu\sigma}-(p_2+p_{V_2})_{\mu}\varepsilon_{V_2\sigma}^*]\mathcal{F}(p_i^2)\\
                &=g_1g_2g_3\int\frac{d^4p_1}{(2\pi)^4 }\frac{(p_D+p_1)_{\nu}\varepsilon_{\alpha\beta\rho\delta}p_2^{\alpha}p_{V_1}^{\beta}\varepsilon_{V_1}^{\delta*}(g^{\mu\nu}-\frac{p_3^{\mu}p_3^{\nu}}{p_3^2})}{(p_1^2-m_{K^*}^2+i\epsilon)(p_2^2-m_K^2+i\epsilon)(p_3^2-m_K^2+i\epsilon)}\\
                &\times[(p_3+p_{V_2})^{\rho}\varepsilon_{V_2\mu}^*+(p_2-p_3)_{\lambda}\varepsilon_{V_2}^{\lambda*}g_{\mu}^{\rho}-(p_2+p_{V_2})_{\mu}\varepsilon_{V_2}^{\rho*}]\mathcal{F}(p_i^2)\\
                &=g_1g_2g_3\int\frac{d^4p_1}{(2\pi)^4 }\frac{4\mathcal{F}(p_i^2)}{p_3^2(p_1^2-m_{K^*}^2+i\epsilon)(p_2^2-m_K^2+i\epsilon)(p_3^2-m_K^2+i\epsilon)}\\
                &\times \{\epsilon_{\alpha\beta\delta\lambda}p_{V_1}^{\alpha}p_1^{\beta}\varepsilon_{V_1}^{\delta*}\varepsilon_{V_2}^{\lambda*}[(p_1\cdot p_{V_2})(p_{V_1}^2+p_{V_1}\cdot p_{V_2}-p_1\cdot p_{V_2})+p_1^2(p_{V_2}^2+p_{V_1}\cdot p_{V_2})\\
                &\ \ \ \ \ \ \ \ \ \ \ \ \ \ \ \ \ \ \ \ \ \ \ \ \ -(p_1\cdot p_{V_1})(p_{V_2}^2+p_{V_1}\cdot p_{V_2}+p_1\cdot p_{V_2})]\\
                &+\epsilon_{\alpha\beta\delta\lambda}p_{V_1}^{\alpha}p_{V_2}^{\beta}\varepsilon_{V_1}^{\delta*}p_1^{\lambda}[p_3^2(p_1\cdot\varepsilon_{V_2}^*)]
                    \}
            \end{split}
        \end{align}

        \item $\tilde{\mathcal{I}}[\text{(PC)},\ K^*,\ \bar{K},\ (K^*)]$
        \begin{align}
            \begin{split}
                i\mathcal{M}
                &=g_1g_2g_3\int\frac{d^4p_1}{(2\pi)^4}\frac{(p_D+p_3)_{\mu}\epsilon_{\alpha\beta\sigma\lambda}p_2^{\alpha}p_{V_2}^{\beta}\varepsilon_{V_2}^{\lambda*}(g^{\mu\nu}-\frac{p_1^{\mu}p_1^{\nu}}{p_1^2})(g^{\rho\sigma}-\frac{p_2^{\rho}p_2^{\sigma}}{p_2^2})}{(p_1^2-m_{K^*}^2+i\epsilon)(p_2^2-m_{K^*}^2+i\epsilon)(p_3^2-m_K^2+i\epsilon)}\\
                &\times[(p_1+p_{V_1})_\rho\varepsilon_{V_1\nu}^*+(p_2-p_{V_1})_{\nu}\varepsilon_{V_1\rho}^*-(p_1+p_2)_{\delta}\varepsilon_{V_1}^{\delta*}g_{\nu\rho}]\mathcal{F}(p_i^2)\\
                &=g_1g_2g_3\int\frac{d^4p_1}{(2\pi)^4}\frac{(p_D+p_3)_{\mu}\epsilon_{\alpha\beta\sigma\lambda}p_2^{\alpha}p_{V_2}^{\beta}\varepsilon_{V_2}^{\lambda*}(g^{\mu\nu}-\frac{p_1^{\mu}p_1^{\nu}}{p_1^2})}{(p_1^2-m_{K^*}^2+i\epsilon)(p_2^2-m_{K^*}^2+i\epsilon)(p_3^2-m_K^2+i\epsilon)}\\
                &\times[(p_1+p_{V_1})^\sigma\varepsilon_{V_1\nu}^*+(p_2-p_{V_1})_{\nu}\varepsilon_{V_1}^{\sigma*}-(p_1+p_2)_{\delta}\varepsilon_{V_1}^{\delta*}g_{\nu}^{\sigma}]\mathcal{F}(p_i^2)\\
                &=g_1g_2g_3\int\frac{d^4p_1}{(2\pi)^4}\frac{4\mathcal{F}(p_i^2)}{p_1^2(p_1^2-m_{K^*}^2+i\epsilon)(p_2^2-m_{K^*}^2+i\epsilon)(p_3^2-m_K^2+i\epsilon)}\\
                &\times\{ \epsilon_{\alpha\beta\delta\lambda} p_{V_1}^{\alpha}p_{V_2}^{\beta}\varepsilon_{V_1}^{\delta*}\varepsilon_{V_2}^{\lambda*}[(p_1\cdot p_{V_1})^2+(p_1\cdot p_{V_1})(p_1\cdot p_{V_2})-p_1^2(p_{V_1}^2+p_{V_1}\cdot p_{V_2})]\\
                &+ \epsilon_{\alpha\beta\delta\lambda} p_{1}^{\alpha}p_{V_2}^{\beta}\varepsilon_{V_1}^{\delta*}\varepsilon_{V_2}^{\lambda*}[-(p_1\cdot p_{V_1})^2-(p_1\cdot p_{V_1})(p_1\cdot p_{V_2})+p_1^2(p_{V_1}^2+p_{V_1}\cdot p_{V_2})]\\
                &+\epsilon_{\alpha\beta\delta\lambda} p_{V_1}^{\alpha}p_{V_2}^{\beta}p_1^{\delta}\varepsilon_{V_2}^{\lambda*}[(p_{V_2}\cdot \varepsilon_{V_1}^*-p_1\cdot \varepsilon_{V_1}^*)p_1^2] \}
            \end{split}
        \end{align}

        \item $\tilde{\mathcal{I}}[\text{(PV)},\ K^*,\ \bar{K}^*,\ (K)]$
        \begin{align}
            \begin{split}
        i\mathcal{M}
        &=g_1g_2g_3\int\frac{d^4p_1}{(2\pi)^4}\frac{\epsilon_{\alpha_1\beta_1\mu\delta}p_1^{\alpha_1}p_{V_1}^{\beta_1}\varepsilon_{V_1}^{\delta*}(g^{\mu\rho}-\frac{p_1^{\mu}p_1^{\rho}}{p_1^2})\epsilon^{\alpha_2\beta_2\nu\lambda}p_{3\alpha_2}p_{V_2\beta_2}\varepsilon_{\phi\lambda}^*(g_{\nu\rho}-\frac{p_{3\nu}p_{3\rho}}{p_3^2})}{(p_1^2-m_{K^*}^2+i\epsilon)(p_2^2-m_K^2+i\epsilon)(p_3^2-m_{K^*}^2+i\epsilon)}\mathcal{F}(p_i^2)\\
        &=g_1g_2g_3\int\frac{d^4p_1}{(2\pi)^4}\frac{\epsilon_{\alpha_1\beta_1\mu\delta}p_1^{\alpha_1}p_{V_1}^{\beta_1}\varepsilon_{V_1}^{\delta*}\times\epsilon^{\alpha_2\beta_2\mu\lambda}p_{3\alpha_2}p_{V_2\beta_2}\varepsilon_{V_2\lambda}^*}{(p_1^2-m_{K^*}^2+i\epsilon)(p_2^2-m_K^2+i\epsilon)(p_3^2-m_{K^*}^2+i\epsilon)}\mathcal{F}(p_i^2)\\
        &=g_1g_2g_3\int\frac{d^4p_1}{(2\pi)^4}\frac{\mathcal{F}(p_i^2)}{(p_1^2-m_{K^*}^2+i\epsilon)(p_2^2-m_K^2+i\epsilon)(p_3^2-m_{K^*}^2+i\epsilon)}\\
        &\times \{(\varepsilon_{V_1}^*\cdot\varepsilon_{V_2}^*)[(p_1^2-p_1\cdot p_{V_1})(p_{V_1}\cdot p_{V_2})+(p_{V_1}^2-p_1\cdot p_{V_1})(p_1\cdot p_{V_2})]\\
        &-(p_1\cdot \varepsilon_{V_1}^*)(p_1\cdot\varepsilon_{V_2}^*)(p_{V_1}\cdot p_{V_2})\\
        &+(p_1\cdot \varepsilon_{V_1}^*)(p_{V_1}\cdot \varepsilon_{V_2}^*)(p_1\cdot p_{V_2})\\
        &+(p_{V_2}\cdot \varepsilon_{V_1}^*)(p_1\cdot \varepsilon_{V_2}^*)[(p_1\cdot p_{V_1})-p_{V_1}^2]\\
        &+(p_{V_2}\cdot \varepsilon_{V_1}^*)(p_{V_1}\cdot \varepsilon_{V_2}^*)[(p_1\cdot p_{V_1})-p_{1}^2]\}
    \end{split}
\end{align}

\item $\tilde{\mathcal{I}}[\text{(PV)},\ K^*,\ \bar{K}^*,\ (K^*)]$
\begin{align}
    \begin{split}
        i\mathcal{M}
        &=g_1g_2g_3\int\frac{d^4p_1}{(2\pi)^4}\frac{(g^{\mu\lambda}-\frac{p_1^{\mu}p_1^{\lambda}}{p_1^2})(g^{\rho\sigma}-\frac{p_2^{\rho}p_2^{\sigma}}{p_2^2})(g_{\nu\lambda}-\frac{p_{3\nu}p_{3\lambda}}{p_3^2})}{(p_1^2-m_{K^*}^2+i\epsilon)(p_2^2-m_{K^*}^2+i\epsilon)(p_3^2-m_{K^*}^2+i\epsilon)}\\
        &\times[(p_1+p_{V_1})_{\rho}\varepsilon_{V_1}^{\delta*}g_{\mu\delta}+(p_2-p_{V_1})_{\mu}\varepsilon_{V_1}^{\delta*}g_{\delta\rho}-(p_1+p_2)_{\delta}\varepsilon_{V_1}^{\delta*}g_{\mu\rho}]\\
        &\times[(p_3+p_{V_2})_{\sigma}\varepsilon_{V_2\beta}^{*}g^{\nu\beta}-(p_2+p_{V_2})^{\nu}\varepsilon_{V_2}^{\beta*}g_{\beta\sigma}+(p_2-p_3)_{\beta}\varepsilon_{V_2}^{\beta*}g_{\sigma}^{\nu}]\mathcal{F}(p_i^2)\\
        &=g_1g_2g_3\int\frac{d^4p_1}{(2\pi)^4}\frac{(p_1^2g^{\mu\lambda}-p_1^{\mu}p_1^{\lambda})(p_2^2g^{\rho\sigma}-p_2^{\rho}p_2^{\sigma})(p_3^2g_{\nu\lambda}-p_{3\nu}p_{3\lambda})}{p_1^2p_2^2p_3^2(p_1^2-m_{K^*}^2+i\epsilon)(p_2^2-m_{K^*}^2+i\epsilon)(p_3^2-m_{K^*}^2+i\epsilon)}\\
        &\times[(p_1+p_{V_1})_{\rho}\varepsilon_{V_1}^{\delta*}g_{\mu\delta}+(p_2-p_{V_1})_{\mu}\varepsilon_{V_1}^{\delta*}g_{\delta\rho}-(p_1+p_2)_{\delta}\varepsilon_{V_1}^{\delta*}g_{\mu\rho}]\\
        &\times[(p_3+p_{V_2})_{\sigma}\varepsilon_{V_2\beta}^{*}g^{\nu\beta}-(p_2+p_{V_2})^{\nu}\varepsilon_{V_2}^{\beta*}g_{\beta\sigma}+(p_2-p_3)_{\beta}\varepsilon_{V_2}^{\beta*}g_{\sigma}^{\nu}]\mathcal{F}(p_i^2)
    \end{split}
\end{align}

\item $\tilde{\mathcal{I}}[\text{(PV)},\ K^*,\ \bar{K}^*,\ (\kappa)]$
\begin{align}
    \begin{split}
        i\mathcal{M}
        &=g_1g_2g_3\int\frac{d^4p_1}{(2\pi)^4}\frac{(g^{\mu\rho}-\frac{p_1^{\mu}p_1^{\rho}}{p_1^2})(g_{\mu\sigma}-\frac{p_{3\mu}p_{3\sigma}}{p_3^2})\varepsilon_{V_1\rho}^*\varepsilon_{V_2}^{\sigma*}}{(p_1^2-m_{K^*}^2+i\epsilon)(p_2^2-m_{\kappa}^2+i\epsilon)(p_3^2-m_{K^*}^2+i\epsilon)}\mathcal{F}(p_i^2)\\
        &=g_1g_2g_3\int\frac{d^4p_1}{(2\pi)^4}\frac{(p_1^2g^{\mu\rho}-p_1^{\mu}p_1^{\rho})(p_3^2g_{\mu\sigma}-p_{3\mu}p_{3\sigma})\varepsilon_{V_1\rho}^*\varepsilon_{V_2}^{\sigma*}}{p_1^2p_3^2(p_1^2-m_{K^*}^2+i\epsilon)(p_2^2-m_{\kappa}^2+i\epsilon)(p_3^2-m_{K^*}^2+i\epsilon)}\mathcal{F}(p_i^2)\\
        &=g_1g_2g_3\int\frac{d^4p_1}{(2\pi)^4}\frac{\mathcal{F}(p_i^2)}{p_1^2p_3^2(p_1^2-m_{K^*}^2+i\epsilon)(p_2^2-m_{\kappa}^2+i\epsilon)(p_3^2-m_{K^*}^2+i\epsilon)}\\
        &\times\{(\varepsilon_{V_1}^*\cdot\varepsilon_{V_2}^*)p_1^2p_3^2\\
        &+(p_1\cdot\varepsilon_{V_1}^*)(p_1\cdot\varepsilon_{V_2}^*)[(p_1\cdot p_3)-(p_{V_1}+p_{V_2})^2]\\
        &+(p_1\cdot\varepsilon_{V_1})(p_{V_1}\cdot\varepsilon_{V_2})[p_1\cdot(p_{V_1}+p_{V_2})]\\
        &+(p_{V_2}\cdot\varepsilon_{V_1})(p_1\cdot\varepsilon_{V_2})p_1^2\\
        &-(p_{V_2}\cdot\varepsilon_{V_1})(p_{V_1}\cdot\varepsilon_{V_2})p_1^2\}
    \end{split}
\end{align}

\item $\tilde{\mathcal{I}}[\text{(PV)},\ K,\ \bar{K},\ (K)]$
\begin{align}
    \begin{split}
        i\mathcal{M}&=g_1g_2g_3\int\frac{d^4p_1}{(2\pi)^4}\frac{(p_1+p_2)_{\mu}(p_2-p_3)_{\nu}\varepsilon_{V_1}^{\mu*}\varepsilon_{V_2}^{\nu*}}{(p_1^2-m_K^2+i\epsilon)(p_2^2-m_K^2+i\epsilon)(p_3^2-m_K^2+i\epsilon)}\mathcal{F}(p_i^2)\\
        &=g_1g_2g_3\int\frac{d^4p_1}{(2\pi)^4}\frac{4\mathcal{F}(p_i^2)}{(p_1^2-m_K^2+i\epsilon)(p_2^2-m_K^2+i\epsilon)(p_3^2-m_K^2+i\epsilon)}\\
        &\times(p_1\cdot\varepsilon_{V_1}^*)(p_1\cdot\varepsilon_{V_2}^*-p_{V_1}\cdot\varepsilon_{V_2}^*)
    \end{split}
\end{align}
\item $\tilde{\mathcal{I}}[\text{PV},\ K,\ \bar{K},\ (K^*)]$
\begin{align}
    \begin{split}
        i\mathcal{M}
        &=g_1g_2g_3\int\frac{d^4p_1}{(2\pi)^4}\frac{\epsilon_{\alpha\beta\mu\lambda}p_2^{\alpha}p_{V_1}^{\beta}\varepsilon_{V_1}^{\lambda*}\epsilon_{\alpha_1\beta_1\nu\delta}p_2^{\alpha_1}p_{V_2}^{\beta_1}\varepsilon_{V_2}^{\delta*}(g^{\mu\nu}-\frac{p_2^\mu p_2^{\nu}}{p_2^2})}{(p_1^2-m_K^2+i\epsilon)(p_2^2-m_{K^*}^2+i\epsilon)(p_3^2-m_K^2+i\epsilon)}\mathcal{F}(p_i^2)\\
        &=g_1g_2g_3\int\frac{d^4p_1}{(2\pi)^4}\frac{\epsilon_{\alpha\beta\mu\lambda}p_2^{\alpha}p_{V_1}^{\beta}\varepsilon_{V_1}^{\lambda*}\epsilon_{\alpha_1\beta_1\mu\delta}p_2^{\alpha_1}p_{V_2}^{\beta_1}\varepsilon_{V_2}^{\delta*}}{(p_1^2-m_K^2+i\epsilon)(p_2^2-m_{K^*}^2+i\epsilon)(p_3^2-m_K^2+i\epsilon)}\mathcal{F}(p_i^2)\\
        &=g_1g_2g_3\int\frac{d^4p_1}{(2\pi)^4}\frac{\mathcal{F}(p_i^2)}{(p_1^2-m_K^2+i\epsilon)(p_2^2-m_{K^*}^2+i\epsilon)(p_3^2-m_K^2+i\epsilon)}\\
        &\times\{(\varepsilon_{V_1}^*\cdot\varepsilon_{V_2}^*)[(p_1\cdot p_{V_2})(p_1\cdot p_{V_1}-p_{V_1}^2)+(p_{V_1}\cdot p_{V_2})(p_1\cdot p_{V_1}-p_1^2)]\\
        &+(p_1\cdot\varepsilon_{V_1}^*)(p_1\cdot\varepsilon_{V_2}^*)(p_{V_1}\cdot p_{V_2})\\
        &-(p_1\cdot\varepsilon_{V_1}^*)(p_{V_1}\cdot\varepsilon_{V_2}^*)(p_{V_1}\cdot p_{V_2})\\
        &+(p_{V_2}\cdot\varepsilon_{V_1}^*)(p_1\cdot\varepsilon_{V_2}^*)[p_{V_1}^2-(p_1\cdot p_{V_1})]\\
        &+(p_{V_2}\cdot\varepsilon_{V_1}^*)(p_{V_1}\cdot\varepsilon_{V_2}^*)[p_{1}^2-(p_1\cdot p_{V_1})]\}
    \end{split}
\end{align}
    \end{itemize}


In the above amplitudes the product of the propagators and the form factor can be expanded as:
\begin{align}
    \begin{split}
        \frac{\mathcal{F}(q_i^2)}{D_1D_2D_3}
        &=\frac{1}{(p_1^2-m_1^2)(p_2^2-m_2^2)(p_3^2-m_3^2)}\left(\frac{m_1^2-\Lambda_1^2}{p_1^2-\Lambda_1^2}\right)\left(\frac{m_2^2-\Lambda_2^2}{p_2^2-\Lambda_2^2}\right)\left(\frac{m_3^2-\Lambda_3^2}{p_3^2-\Lambda_3^2}\right)\\
        &=\left(\frac{1}{p_1^2-m_1^2}-\frac{1}{p_1^2-\Lambda_1^2}\right)\left(\frac{1}{p_2^2-m_2^2}-\frac{1}{p_2^2-\Lambda_2^2}\right)\left(\frac{1}{p_3^2-m_3^2}-\frac{1}{p_3^2-\Lambda_3^2}\right)\\
        &=\frac{1}{(p_1^2-m_1^2)(p_2^2-m_2^2)(p_3^2-m_3^2)}-\frac{1}{(p_1^2-\Lambda_1^2)(p_2^2-m_2^2)(p_3^2-m_3^2)}-\frac{1}{(p_1^2-m_1^2)(p_2^2-\Lambda_2^2)(p_3^2-m_3^2)}\\
        &-\frac{1}{(p_1^2-m_1^2)(p_2^2-m_2^2)(p_3^2-\Lambda_3^2)}+\frac{1}{(p_1^2-\Lambda_1^2)(p_2^2-\Lambda_2^2)(p_3^2-m_3^2)}+\frac{1}{(p_1^2-\Lambda_1^2)(p_2^2-m_2^2)(p_3^2-\Lambda_3^2)}\\
        &+\frac{1}{(p_1^2-m_1^2)(p_2^2-\Lambda_2^2)(p_3^2-\Lambda_3^2)}-\frac{1}{(p_1^2-\Lambda_1^2)(p_2^2-\Lambda_2^2)(p_3^2-\Lambda_3^2)} 
    \end{split}
\end{align}


\begin{thebibliography}{99}

\bibitem{BaBar:2003zor}
B.~Aubert \textit{et al.} [BaBar],
Phys. Rev. Lett. \textbf{91} (2003), 171802
doi:10.1103/PhysRevLett.91.171802
[arXiv:hep-ex/0307026 [hep-ex]].

\bibitem{BaBar:2003rdc}
B.~Aubert \textit{et al.} [BaBar],
Phys. Rev. D \textbf{69} (2004), 031102
doi:10.1103/PhysRevD.69.031102
[arXiv:hep-ex/0311017 [hep-ex]].

\bibitem{Belle:2003lsm}
J.~Zhang \textit{et al.} [Belle],
Phys. Rev. Lett. \textbf{91} (2003), 221801
doi:10.1103/PhysRevLett.91.221801
[arXiv:hep-ex/0306007 [hep-ex]].

\bibitem{Brodsky:1981kj}
S.~J.~Brodsky and G.~P.~Lepage,
Phys. Rev. D \textbf{24} (1981), 2848
doi:10.1103/PhysRevD.24.2848

\bibitem{Chernyak:1981zz}
V.~L.~Chernyak and A.~R.~Zhitnitsky,
Nucl. Phys. B \textbf{201} (1982), 492
[erratum: Nucl. Phys. B \textbf{214} (1983), 547]
doi:10.1016/0550-3213(83)90251-1

\bibitem{Chernyak:1983ej}
V.~L.~Chernyak and A.~R.~Zhitnitsky,
Phys. Rept. \textbf{112} (1984), 173
doi:10.1016/0370-1573(84)90126-1

\bibitem{Belle:2003ike}
K.~F.~Chen \textit{et al.} [Belle],
Phys. Rev. Lett. \textbf{91} (2003), 201801
doi:10.1103/PhysRevLett.91.201801
[arXiv:hep-ex/0307014 [hep-ex]].

\bibitem{Kamal:1990ky}
A.~N.~Kamal, R.~C.~Verma and N.~Sinha,
Phys. Rev. D \textbf{43} (1991), 843-854
doi:10.1103/PhysRevD.43.843

\bibitem{Hinchliffe:1995hz}
I.~Hinchliffe and T.~A.~Kaeding,
Phys. Rev. D \textbf{54} (1996), 914-928
doi:10.1103/PhysRevD.54.914
[arXiv:hep-ph/9502275 [hep-ph]].

\bibitem{Bedaque:1993fb}
P.~F.~Bedaque, A.~K.~Das and V.~S.~Mathur,
Phys. Rev. D \textbf{49} (1994), 269-274
doi:10.1103/PhysRevD.49.269
[arXiv:hep-ph/9307296 [hep-ph]].

\bibitem{Bauer:1986bm}
M.~Bauer, B.~Stech and M.~Wirbel,
Z. Phys. C \textbf{34} (1987), 103
doi:10.1007/BF01561122

\bibitem{Cheng:2010rv}
H.~Y.~Cheng and C.~W.~Chiang,
Phys. Rev. D \textbf{81} (2010), 114020
doi:10.1103/PhysRevD.81.114020
[arXiv:1005.1106 [hep-ph]].

\bibitem{Jiang:2017zwr}
H.~Y.~Jiang, F.~S.~Yu, Q.~Qin, H.~n.~Li and C.~D.~L\"u,
Chin. Phys. C \textbf{42} (2018) no.6, 063101
doi:10.1088/1674-1137/42/6/063101
[arXiv:1705.07335 [hep-ph]].

\bibitem{Uppal:1992se}
T.~Uppal and R.~C.~Verma,
Z. Phys. C \textbf{56} (1992), 273-277
doi:10.1007/BF01555524

\bibitem{Bajc:1997ey}
B.~Bajc, S.~Fajfer, R.~J.~Oakes and S.~Prelovsek,
Phys. Rev. D \textbf{56} (1997), 7207-7215
doi:10.1103/PhysRevD.56.7207
[arXiv:hep-ph/9706223 [hep-ph]].

\bibitem{ElHassanElAaoud:1999min}
E.~H.~E.~Aaoud and A.~N.~Kamal,
Phys. Rev. D \textbf{59} (1999), 114013
doi:10.1103/PhysRevD.59.114013
[arXiv:hep-ph/9910350 [hep-ph]].

\bibitem{Hiller:2013cza}
G.~Hiller and R.~Zwicky,
JHEP \textbf{03} (2014), 042
doi:10.1007/JHEP03(2014)042
[arXiv:1312.1923 [hep-ph]].

\bibitem{MARK-III:1991fvi}
D.~Coffman \textit{et al.} [MARK-III],
Phys. Rev. D \textbf{45} (1992), 2196-2211
doi:10.1103/PhysRevD.45.2196

\bibitem{FOCUS:2007ern}
J.~M.~Link \textit{et al.} [FOCUS],
Phys. Rev. D \textbf{75} (2007), 052003
doi:10.1103/PhysRevD.75.052003
[arXiv:hep-ex/0701001 [hep-ex]].

\bibitem{BESIII:2021raf}
M.~Ablikim \textit{et al.} [BESIII],
Phys. Rev. Lett. \textbf{128} (2022) no.1, 011803
doi:10.1103/PhysRevLett.128.011803
[arXiv:2108.02405 [hep-ex]].

\bibitem{dArgent:2017gzv}
P.~d'Argent, N.~Skidmore, J.~Benton, J.~Dalseno, E.~Gersabeck, S.~Harnew, P.~Naik, C.~Prouve and J.~Rademacker,
JHEP \textbf{05} (2017), 143
doi:10.1007/JHEP05(2017)143
[arXiv:1703.08505 [hep-ex]].

\bibitem{LHCb:2018mzv}
R.~Aaij \textit{et al.} [LHCb],
JHEP \textbf{02} (2019), 126
doi:10.1007/JHEP02(2019)126
[arXiv:1811.08304 [hep-ex]].

\bibitem{LeYaouanc:1988fx}
A.~Le Yaouanc, L.~Oliver, O.~Pene and J.~C.~Raynal,

\bibitem{Richard:2016hac}
J.~M.~Richard, Q.~Wang and Q.~Zhao,
[arXiv:1604.04208 [nucl-th]].

\bibitem{Niu:2020gjw}
P.~Y.~Niu, J.~M.~Richard, Q.~Wang and Q.~Zhao,
Phys. Rev. D \textbf{102} (2020) no.7, 073005
doi:10.1103/PhysRevD.102.073005
[arXiv:2003.09323 [hep-ph]].

\bibitem{Kokoski:1985is}
R.~Kokoski and N.~Isgur,
Phys. Rev. D \textbf{35} (1987), 907
doi:10.1103/PhysRevD.35.907

\bibitem{Godfrey:1985xj}
S.~Godfrey and N.~Isgur,
Phys. Rev. D \textbf{32} (1985), 189-231
doi:10.1103/PhysRevD.32.189

\bibitem{Godfrey:1986wj}
S.~Godfrey and R.~Kokoski,
Phys. Rev. D \textbf{43} (1991), 1679-1687
doi:10.1103/PhysRevD.43.1679

\bibitem{Niu:2021qcc}
P.~Y.~Niu, Q.~Wang and Q.~Zhao,
Phys. Lett. B \textbf{826} (2022), 136916
doi:10.1016/j.physletb.2022.136916
[arXiv:2111.14111 [hep-ph]].

\bibitem{Cheng:2021nal}
Y.~Cheng and Q.~Zhao,
Phys. Rev. D \textbf{105} (2022) no.7, 076023
doi:10.1103/PhysRevD.105.076023
[arXiv:2106.12483 [hep-ph]].

\bibitem{Cheng:2023lov}
Y.~Cheng, L.~Qiu and Q.~Zhao,
[arXiv:2302.01210 [hep-ph]].

\bibitem{Guo:2010ak}
F.~K.~Guo, C.~Hanhart, G.~Li, U.~G.~Meissner and Q.~Zhao,
Phys. Rev. D \textbf{83} (2011), 034013
doi:10.1103/PhysRevD.83.034013
[arXiv:1008.3632 [hep-ph]].

\bibitem{ARGUS:1992gpk}
H.~Albrecht \textit{et al.} [ARGUS],
Z. Phys. C \textbf{56} (1992), 7-14
doi:10.1007/BF01589701

\bibitem{BESIII:2017jyh}
M.~Ablikim \textit{et al.} [BESIII],
Phys. Rev. D \textbf{95} (2017) no.7, 072010
doi:10.1103/PhysRevD.95.072010
[arXiv:1701.08591 [hep-ex]].



\end{thebibliography}

\end{document}